# Repeated Growth and Hot Delamination of Single-Crystal Graphene: A Two-Kilometer Process-Design Assessment with hBN as a Separate Case

*Rodney S. Ruoff*[*]

R. S. Ruoff, Center for Multidimensional Carbon Materials (CMCM), Institute for Basic Science (IBS), Ulsan 44919, Republic of Korea

R. S. Ruoff, Department of Chemistry, UNIST, Ulsan 44919, Republic of Korea

R. S. Ruoff, Department of Materials Science and Engineering, UNIST, Ulsan 44919, Republic of Korea

R. S. Ruoff, School of Energy and Chemical Engineering, UNIST, Ulsan 44919, Republic of Korea

E-mail: ruoff@unist.ac.kr

*Funding: This work was supported by the Institute for Basic Science (IBS-R019-D1), Republic of Korea.*

Large single crystals of graphene and hexagonal boron nitride (hBN) remain difficult to manufacture because batch growth and transfer limit area, throughput, and quality. This Perspective asks whether an atomically thin film could instead be grown, hot-delaminated, and regrown repeatedly on a long, reusable single-crystal metal surface. No integrated process is demonstrated. Once rapid full-area growth is available, hot film removal is likely to become the rate-limiting step, whereas usable single-crystal growth area sets the material produced per cycle. Two experiments determine whether the concept merits further development: rapid, damage-free delamination at or near the growth temperature, and acceptable film growth through repeated complete growth-dwell-peel-regrowth cycles. The analysis then examines the crystalline growth surface, layer control, interface and gas chemistries, product capture, quality criteria, energy, and scale. These provisional analyses expose known requirements, define early stop criteria, and guide experiments; they cannot anticipate every coupled constraint or failure mode. Graphene is the quantitative baseline. hBN requires its own M(111), product thickness, apparatus, and scaling strategy. Pilot-scale development is justified only if the two central experiments succeed and the remaining gates can then be refined experimentally.

## 1. Introduction

Large single crystals of graphene and hexagonal boron nitride (hBN) are needed when grain boundaries and other defects would limit electronic, thermal, mechanical, or photonic performance. Chemical vapor deposition (CVD) can grow these films on Cu, Ni, and related metals. Our 2009 demonstration of centimeter-scale, predominantly monolayer graphene by CVD on Cu helped establish metal-catalyzed CVD as a practical route toward scaled graphene production.[1] In that study, however, the Cu was dissolved and polymer support was needed to reduce transfer damage. Scaled growth therefore did not solve continuous removal from a reusable substrate. Most wet and dry transfer methods are batch processes and can tear, wrinkle, or contaminate the film. Once full-area growth is rapid, removal and immediate product capture are likely to set the cycle rate. The amount produced in one cycle is limited principally by the usable single-crystal growth area; the delivered rate is that accepted area divided by the complete cycle time. The central manufacturing problem is therefore not growth alone. It is coupling growth to removal while preserving crystallinity over useful areas and at useful throughput.

I examine a deliberately extreme architecture: simultaneous growth of single-crystal graphene (SCG) on a fixed, flat, 2 km × 30 cm M(111) surface, followed by cyclic in situ delamination and winding. Here, M denotes a metal and (111) its close-packed crystallographic surface. The M(111) surface could be a converted metal foil or an epitaxial metal film on a long template. The 2 km length is a stress test, not a minimum requirement: an extreme case exposes scaling limits that a modest one can hide. Graphene is the quantitative baseline. I treat single-crystal hBN (SC-hBN) as a separate case because its growth chemistry, product thickness, release mechanics, apparatus, and likely scaling strategy need not inherit those of graphene. My aims are to ask whether the architecture is internally consistent, identify which elements have precedent, and define experiments that could disprove it. I do not present a build-ready machine, and I do not claim that every requirement or failure mode can be anticipated before the first coupled experiments.

Two possible uses show why length and area matter. Long SCG ribbons could provide high specific strength at very low mass.[2] Large, low-defect graphene is also being developed for low-power integrated photonics and data movement.[3] Commercial investment in graphene photonics signals interest in large-area material,[4] but not manufacturing feasibility. Relevant growth conditions already span an unusually broad range: radical-mediated graphene growth on Cu(111) has been reported at room temperature, whereas classical studies established temperature-dependent carbon surface phases on Ni(111).[5–7] Neither result addresses hot cyclic removal. Both prospective uses require areas or lengths that batch growth and transfer do not readily provide.

Growth must therefore be coupled to removal so that the exposed surface can regrow. Can an atomically thin single crystal be delaminated cleanly and rapidly at, or near, the growth temperature, and can an accepted film then form again during each cycle? Those are the decisive questions. I define the experiments needed to answer them.

This distinction sets the scope of the Perspective. Hot peeling and repeated complete-cycle growth are the critical experiments; if either fails without a practical remedy, the proposed architecture fails. The sections that follow ask what else scaled production could require, including a crystalline substrate, growth and interface chemistry, carrier capture, product acceptance, atmosphere control, energy, and module length. They are an informed first pass, not a claim that the engineering has been solved. A process that has not yet been attempted cannot be specified perfectly from literature and calculations alone. I expect the first coupled experiments to expose interactions and failure modes that cannot now be anticipated, and the design gates should be revised when they do.

## 2. Process architecture: a cyclic traveling peel front

The architecture must be qualified as a complete cycle. Work in progress in our group has yielded full-area graphene coverage under the conditions considered here, with $t_g$ = 180 s[8]. Because the result is not yet publicly available, I use 180 s as a provisional process input rather than as an established performance benchmark. At $v_p = v_r$ = 1 m s$^{-1}$, a head takes ~2000 s to peel from 0 to 2 km and another ~2000 s to return empty. Allowing 200 s for offloading, reloading, and turnaround gives ~4200 s. With these illustrative inputs, removal and the associated carriage cycle—not the 180 s growth interval—set the gross rate. About half of the 4200 s arises from the empty return in this particular one-carriage layout, so 4200 s is not an intrinsic peel-time limit. A film that reaches coverage in 180 s would remain under the selected process conditions for ~4020 s before the next scheduled peel. I do not require a separate abstract "stability" measurement if repeated full cycles yield films that meet the same layer, crystallinity, and defect criteria at removal. Repeated growth, dwell, peel, and regrowth is the direct test. Local feed gating, a changed background atmosphere, or a shorter module should be tested only if product quality does not remain acceptable through that cycle.

The baseline is a fixed, straight 2 km × 30 cm candidate M(111) track operated by one traveling carriage (Figure 1a). To complete the single-pass material accounting, I use one notional carriage arrangement: at 0 km, the carriage receives a high-temperature carrier-supply reel and an empty product mandrel. It then moves to 2 km while peeling, capturing the film on a carrier at or near the crack front, cooling the supported product over a short carriage-mounted path, and winding it on a carriage-mounted take-up. At 2 km, the ~600 $m^2$ product roll is offloaded; the carriage lifts, returns empty to 0 km, and is reloaded. Behind the moving peel front, the M(111) surface is exposed and allowed to regrow. Acceptance of the next film after the complete cycle determines whether that operating condition survives. The 200 s end-handling allowance is shared between the two ends. This completes the single-head material balance without a 2 km unsupported film span; it does not establish the final pickup, supported-cooling, or winding hardware. A flat in-plane racetrack is excluded because a wide single-crystal ribbon can bend around an axis parallel to its width but cannot turn within its plane without stretching, wrinkling, or tearing. Any loop must therefore curve out of plane.

Gross geometric capacity could be doubled by stacking two fixed flat tracks within one enclosure (Figure 1b1). An out-of-plane band loop with pickup at both ends is more continuous but would subject a single-crystal metal belt to repeated bending and fatigue (Figure 1b2). I therefore retain the fixed linear track as the baseline and the stacked-track arrangement as the lower-risk scale-up option.

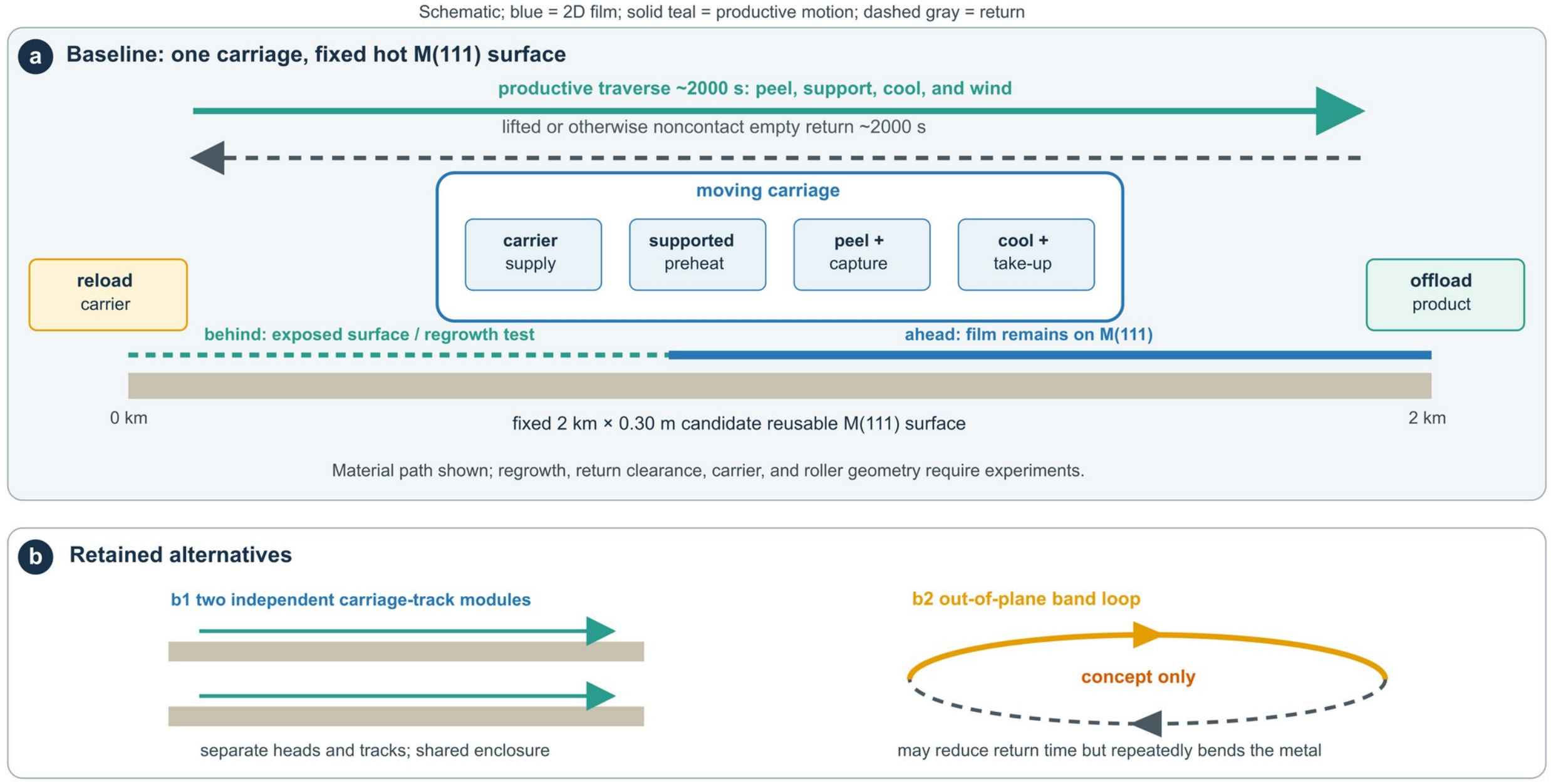

Figure 1. Schematic single-head material-handling cycle and retained alternatives; dimensions other than the stated track length and width are not to scale. M(111) denotes the (111) surface of the selected metal. (a) One carriage moves from 0 to 2 km while peeling and capturing the film on a carrier, cooling it while supported, and winding it; the carriage then returns lifted or otherwise without contacting the regrown film. The fixed 2 km × 0.30 m candidate M(111) surface remains hot. Offloading and reloading share a total 200 s end-handling allowance, giving an illustrative 4200 s cycle. Blue denotes film that remains on M(111) ahead of the moving peel front; the dashed teal line marks the exposed surface and regrowth test behind it. The solid teal arrow denotes productive motion and the dashed gray arrow denotes empty return. The drawing completes single-pass material accounting but does not select carrier or roller geometry. (b) Two independent carriage-track modules could share an enclosure. An out-of-plane loop could reduce idle return but would cyclically bend the metal; it is retained only as a concept.

The 2 km length is not fundamental. The same growth-delamination sequence could be tested on meter-scale tracks or tracks tens of meters long. At smaller scale, the substrate could translate or rotate past a fixed delamination station. The decisive interface physics is unchanged; handling, throughput, energy use, and failure exposure change with length.

The kilometer-scale substrate must remain hot rather than undergo repeated bulk thermal cycles. Heating 2 km of Ni through $\Delta T \approx 1000$ °C produces ~26 m of expansion ($\alpha Ni \approx 13 \times 10^{-6}$ $K^{-1}$); an epitaxial Ni film on sapphire ($\alpha \approx 8 \times 10^{-6}$ $K^{-1}$) would also accumulate large differential strain. Sliding mounts and an overlength enclosure would have to accommodate the one-time expansion. A cooler peel zone or film-side cooling would avoid bulk cycling but impose repeated local thermal gradients. Isothermal operation also avoids cooling-induced carbon precipitation, although the required mechanical design remains unproven.

## 3. Substrate selection

For the nominal ~1050 °C graphene process, I prefer Ni-rich Ni–Cu(111), retain pure Ni(111) as a fallback, and reject Cu(111).

Cu(111) is unsuitable for the nominal hot process. At 1050 °C, Cu is at 0.97 of its absolute melting temperature (1358 K), so grooving, faceting, sublimation, and dewetting of a Cu film become serious risks. Its low carbon solubility remains useful at lower temperature. Thus, single-crystal Cu(111) or Cu–Ni(111) on c-plane sapphire is a viable option for the reported room-temperature radical-mediated graphene process,[5] but that result does not establish an hBN route.

Pure Ni(111) remains viable at 1050 °C (0.77 of its absolute melting temperature). Its surface is mobile enough for step-flow smoothing while retaining thermal margin. High carbon solubility can produce multilayer graphene during cooling, but continuous isothermal operation should reduce that precipitation pathway. The dissolved-carbon inventory may then buffer surface carbon activity; this benefit must be demonstrated over repeated cycles.

Ni-rich Ni–Cu(111), or possibly Ni–Co(111), is preferred because alloying can reduce carbon solubility while retaining Ni's catalytic activity and ~1.2% lattice match to graphene. I propose ~Ni80Cu20(111) only as a starting composition; alloy composition is a process variable, not an established optimum. Preferential Cu evaporation at 1050 °C would shift the surface composition during long runs, so Cu activity would require monitoring and replenishment. Ni–Co avoids Cu loss but needs its own layer-control map.

## 4. Growth chemistry

For graphene, dilute $CH_4$ in $H_2$ supplies surface carbon while the metal contains a controlled dissolved-carbon inventory. On Ni(111), carbon activity and temperature determine transitions from a dilute surface phase to a condensed graphitic monolayer and then multilayer graphite.[6, 7] Isothermal operation within the monolayer window could maintain $n \approx 1$ without cooling-induced precipitation, but that window has not been mapped through repeated peel-regrow cycles. Lean feed is needed because gas-phase pyrolysis would deposit amorphous carbon and create pinning sites. It is well established that graphene need not grow from one nucleus to form a large single crystal. On suitable Cu(111) and related surfaces, many nuclei can share one orientation and merge without a detectable grain boundary by the methods used.[9–11] The proposed process therefore does not require a retained seed after each peel. It requires this nucleation, alignment, and essentially seamless merging to repeat after removal, without accumulating layer errors or other defects. Reference 8 identifies work in progress in our group that has yielded a ~180 s full-coverage time under the conditions considered here. Because the result is not yet publicly available, I use 180 s as a provisional design input rather than as an established performance benchmark. The public report should state the substrate composition and orientation, dimensions, temperature, pressure, gas chemistry, layer number, whether the result is initial growth or regrowth, and the coverage definition and mapping. A full-cycle test then lets the film grow, remain under the selected conditions until the scheduled peel, undergo removal, and regrow repeatedly. If each product meets the same declared acceptance criteria, that result is sufficient evidence of operational cycle stability. Feed-on/feed-off studies are diagnostic tests if the repeated cycle fails or if local gating is being considered; they are not an additional prerequisite. Published growth times span 30 s for fold-free bilayer graphene on a 2-inch Cu–Ni(111)/sapphire film,[12] 20 min for meter-scale SCG on Cu(111),[9] and ~4 h for continuous wafer-scale graphene by room-temperature radical-mediated growth.[5] Related 2025 studies reported edge-fed layer-controlled graphene and roll-to-roll fractional-layer graphene[13, 14], but neither tested repeated hot release. These are not “apples-to-apples” manufacturing comparisons because product, area, transfer, and acceptance definitions differ. None demonstrates the proposed repeated hot-release cycle.

SC-hBN requires its own manufacturing choices. Cu–Ni(111), including $Cu_{0.8}Ni_{0.2}$(111)/sapphire, is an important candidate for monolayer hBN, whereas Ni(111) has produced trilayer and substantially thicker single-crystal films.[15–19] The latter thickness can be a significant handling and application advantage. M(111) composition, orientation, temperature, B and N activities, and target thickness must therefore be tried and perfected for the intended hBN product; there is no reason to assume the graphene optimum. hBN also has distinct B and N sublattices, so rotational variants, 180° inversion domains, and antiphase boundaries must be mapped. Many aligned nuclei may merge, as for graphene, but the binary lattice creates additional ways for a boundary to remain. The proposed seeded superstructure strategy is computational guidance, not a production demonstration.[20]

Boron uptake creates an additional hazard on Ni. The Ni–$Ni_3B$ eutectic is ~1093 °C,[21] only ~40 °C above the nominal graphene set point, and nickel borides form readily. SC-hBN growth on Ni must therefore remain surface-limited, with low B and N activities and minimal bulk B uptake. A lower temperature, initially ~950–1000 °C, would increase the margin, but the appropriate catalyst and window require a separate Ni–B–N phase-chemistry map.

Recent studies provide relevant but different hBN precedents. At centimeter scale, Oh et al. grew an approximately 3-nm-thick epitaxial hBN film on bulk Ni(111) and repeated growth and electrochemical bubbling transfer ten times.[15] They reported that the Ni(111) orientation, surface terraces, and roughness remained essentially unchanged. Single-crystal monolayer hBN has been grown across a 5.08-cm Cu(111)/sapphire wafer,[16] and wafer-scale single-crystal trilayer hBN has been grown on Ni(111).[17] Wang et al. reported ~10 min growth of a 10.16-cm single-crystal hBN monolayer on $Cu_{0.8}Ni_{0.2}$(111)/sapphire; their transfer etched the alloy film.[18] Yang et al. reported a 3–30-nm-thick, 5.08-cm single-crystal rhombohedral BN (rBN) film by solid-liquid-interface-mediated epitaxy on Ni(111)/sapphire. They proposed that Si creates a thin Ni–Si surface liquid layer that stores B and N while the underlying solid Ni(111) preserves epitaxial orientation.[19] A second preprint reports wafer-scale, 1–8-layer single-crystal hBN by deposit-then-crystallize growth on Cu–Ni(111).[22] There is no published evidence for rapid hot release of these hBN films. That absence is the reason to perform the experiment, not a reason to exclude hBN. I also do not assume that graphene and hBN share one apparatus. Further studies will likely lead to different release, carrier, module-length, and scaling strategies for monolayer and thicker hBN products.

A non-epitaxial alternative is carbon precipitation from molten Ni–C, for which supersaturation and cooling rate control the segregated carbon.[23] This has been demonstrated for bulk graphite morphologies, not planar SCG. Related solubility-control precedents exist in solid Cu–Ni and Ni–Mo.[24, 25] I assess liquid-mediated growth briefly in Section 8.3 and in greater detail in Supporting Information Section S7.

## 5. The graphene-M(111) interface: adsorption and effective adhesion

The graphene-Ni(111) interface controls both epitaxial registry and delamination. It cannot be described adequately as either strongly covalent or purely physisorptive. High-level calculations show orbital hybridization at the short equilibrium distance, but an adsorption *energy* of physisorption magnitude.

Bond character, adsorption *energy*, equilibrium distance, and effective fracture *energy* must therefore be treated as distinct quantities.

Random-phase-approximation (RPA/ACFDT) calculations find $\pi$–d($z^2$) hybridization at 2.17 Å and an adsorption energy of only ~67 meV per carbon (~0.4 J $m^{-2}$), an order of magnitude below covalent chemisorption (0.5–2 eV).[26] The same calculations identify a second, purely physisorptive minimum near 3.3 Å that is only a few meV higher.[26, 27] Independent RPA work frames Ni(111) as a "delicate competition" between a weak chemisorption well (~2.3 Å) and a physisorption well (~3.25 Å);[27] Wannier- and dispersion-based treatments reproduce the same ~2.1 Å geometry with physisorption-scale energetics.[28] A full method-by-method compilation of computed and measured values, with the unit conversions, is given in the Supporting Information (Table S1).[S1–S7]

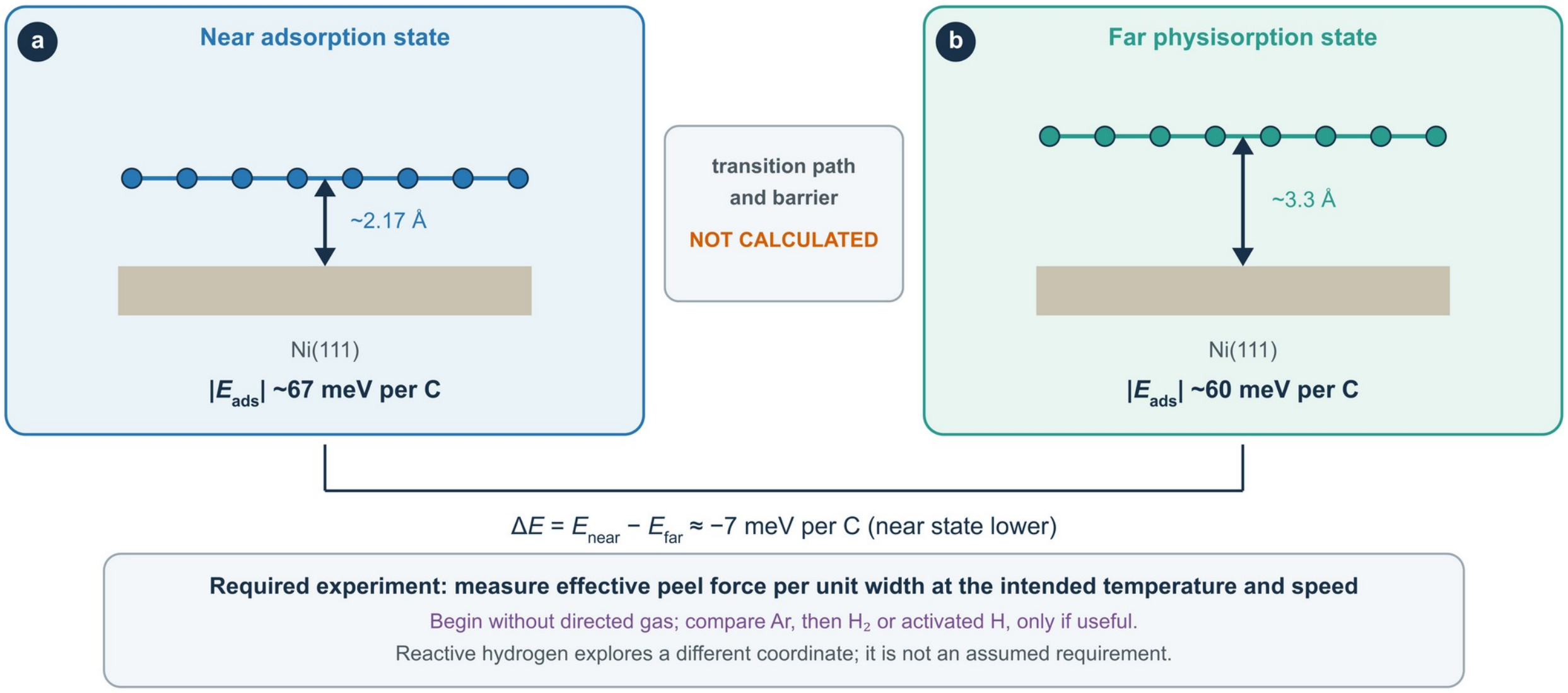

Figure 2. Schematic of calculated clean graphene/Ni(111) adsorption states; distances are not to scale, and the horizontal arrangement does not represent a transition coordinate. RPA calculations give a near state at ~2.17 Å and $|E_{ads}|$ ~67 meV per C and a far physisorption state at ~3.3 Å and $|E_{ads}|$ ~60 meV per C.[26, 27] Here, $|E_{ads}|$ is the binding-energy magnitude. The relation $\Delta E = E_{near} - E_{far} \approx -7$ meV per C places the near state lower in calculated energy, but this difference is not a transition barrier. Neither a connecting path nor a finite-temperature peel free energy has been calculated. The required experiment is a measurement of effective peel force per width at the intended temperature and speed. $H_2$ or activated H introduces a different chemical coordinate and is an optional comparison, not an assumed requirement.

I emphasize that the intrinsic work of separation (~0.4 J $m^{-2}$) is a lower bound, not the peel energy. Mechanical measurements include rate-dependent dissipation, metal plasticity, roughness, pinning, mode mixity, and contamination. Reported *effective* adhesion values span ~0.7–6.8 J $m^{-2}$, including ~6.8 J $m^{-2}$ from a StereoDIC blister test on Ni (Supporting Information, Section S2).[29] The relevant design input is therefore the *effective* adhesion energy measured at the intended speed, temperature, atmosphere, and interface condition.

The calculated near and far minima differ by only a few meV per C, but that energy difference is not the transition barrier. Hydrogenation can shift graphene/Ni toward a more weakly bound state,[30] which justifies testing $H_2$ but does not show that intercalation is required. The first experiment should measure mechanical peeling with no directed gas at the crack tip, using only the selected background growth atmosphere. Local Ar flow can then test pressure, transport, and heat-transfer effects without intentionally changing the interface chemistry. $H_2$ or activated H should be compared as a reactive condition. If a gas is delivered into the opened interface, it needs to advance only a short distance ahead of the mechanical front; it is not intended to penetrate the basal plane. None of these conditions has demonstrated 30 cm-wide peeling at 1 m $s^{-1}$ and high temperature. Effective adhesion may remain several J $m^{-2}$ and increase with speed.

Temperature may make release harder or easier through competing effects. Thermal expansion of Ni *stabilizes the* near *chemisorption well*,[S4] and H-intercalated graphene on SiC de-intercalates near ~900 °C,[31] both unfavorable at 1050 °C. Thermal fluctuations can lower an effective fracture threshold,[32] but this equilibrium result is not a peel-rate law. The net behavior at *the* proposed temperature is therefore unknown and must be measured directly.

## 6. Peeling and delamination mechanics

The design requires strong registry during growth and weak adhesion during removal. It does not require chemical intercalation as an assumption. I propose a controlled sequence: (1) peel with no directed crack-tip gas; (2) repeat with local Ar; and (3) repeat with $H_2$, activated H, or another candidate reactant only if a reactive condition is useful. Hydrogen intercalation produces quasi-free-standing graphene on SiC,[31] and hydrogenation can weaken graphene-metal coupling,[30] but neither result demonstrates hot peeling from Ni. For every condition, measure effective adhesion, front stability, tears, residue, edge damage, and the quality of the subsequently regrown film. A sequence of accepted products from repeated complete cycles is the relevant test. Separate feed-on/feed-off stability measurements become useful if that sequence fails or if gating is being designed. Directed Ar might assist by gas pressure or interface flushing, or it might destabilize the film; no directed gas may prove best. The intercalated SiC state reverses below the proposed 1050 °C set point,[31] so $H_2$ is an experimental variable, not the baseline solution.

Temperature should be varied independently of the local peel-zone atmosphere. In the isothermal mode, growth and peeling occur at one temperature. If a reactive gas is used, any weakly bound state must form faster than it desorbs. In the two-temperature mode, growth occurs hot and peeling occurs in a cooler local zone. The latter adds thermal-mismatch and precipitation risks but may stabilize a decoupled interface. No directed gas, Ar, and reactive-gas conditions should first be compared isothermally; a cooler zone should be added only if the results justify it. The preferred atmosphere and peel temperature are experimental outcomes, not design inputs.

Peeling also requires control of the fracture path. A seed tab or carrier must initiate separation, and the crack follows the weakest interface. Graphite interlayer cohesion (~0.29 J $m^{-2}$) and the far graphene/Ni well (~0.37 J $m^{-2}$) are both below some measured effective metal-interface energies. An $n$-layer product could therefore split within the stack and leave layers on the metal. Carrier/film, interlayer, and film/metal energies must be measured under the same temperature, speed, and atmosphere. The required hierarchy is carrier/film > interlayer > film/metal. This favors carrier-assisted capture at the crack front.

For steady peeling at angle $\theta$, the ideal inextensible-film screen is $F/w \approx G/(1 - \cos\theta)$. At 90° and $G \approx 0.37$ J $m^{-2}$, it gives ~0.4 N $m^{-1}$, or ~0.1 N across 0.30 m. This is an idealized lower-bound screen, not a strength margin. Once the carrier is attached, the peel arm is a hot composite, and measured force can include carrier bending, creep or plastic work, roller friction, and differential contraction. Higher effective adhesion could also raise the interfacial contribution to several N $m^{-1}$. Mean force is then less important than local pinning: adhered patches, step bunches, particles, front skew, and carrier nonuniformity can concentrate stress and nucleate tears. A straight front, a short free span, and a roller radius selected from measurements should reduce this risk.

A peel speed of 1 m $s^{-1}$ is aspirational. Reported graphene and hBN transfer occurs at much lower speeds, including centimeter-per-second roll-to-roll peeling,[S12] sub-millimeter-per-second rate-dependent delamination,[33] and quasi-static blister tests.[29] Tape peeling spans the meter-per-second range but exhibits stick-slip between ~0.25 and 2.45 m $s^{-1}$,[34–36] which warns that a brittle 2D film may tear periodically. Flexible-film and 2D-material fracture models provide useful scaling,[37–40] but do not establish full-width hot peeling under no-flow, Ar, or reactive-gas conditions. The assumed speed must therefore be established through a width-speed-temperature-atmosphere ladder.

The decisive experiment is instrumented peeling of a single-crystal film across increasing widths and speeds at room and elevated temperature, with synchronized force measurement and high-speed imaging. The local atmosphere should be varied from no directed gas to Ar and then to $H_2$ or another reactive candidate. Multiscale modeling, from molecular dynamics to continuum mechanics, could test how temperature, surface reconstruction, gas pressure, optional intercalation, and defect populations affect millisecond-scale fracture.

## 7. Pickup and rollup: functional requirements

No existing device combines full-width hot peeling, immediate product capture, supported cooling, take-up, and regrowth on the exposed metal. I do not prescribe its final mechanical design. Any implementation must (1) maintain a straight peel front; (2) capture the film close enough to that front to avoid a long unsupported hot span; (3) control tension and cooling without introducing wrinkles or tears; (4) collect the product without restacking or contamination; and (5) leave the exposed metal ready to regrow. The candidate test arrangements and mechanical screens in Supporting Information Section S4 and Figures S1 and S2 make these functions explicit. Their rollers, carrier, shroud, nip, cooling path, and take-up are placeholders.

### 7.1 Product capture and mechanical qualification

The pickup system is part of the central experiment because a film is not successfully removed if it cannot be captured intact. A seed tab or carrier must initiate separation, and the crack must remain at the film-metal interface. The first trials should use a short supported span and one carrier close to the crack front. Carrier composition and thickness, roller radius, line tension, temperature, cooling rate, adhesion hierarchy, downstream release, contamination, and reuse should then be varied. No carrier or roller geometry is selected here.

Mechanical qualification cannot be reduced to a nominal roller radius or mean peel force. Section 6 identifies the principal variables; Supporting Information Section S4 preserves the force, heating, and defect-statistics screens. The assumed 1 m $s^{-1}$ speed is aspirational. Qualification should proceed from coupons through increasing speed and width while measuring force, front stability, tears, residue, carrier behavior, and the subsequently regrown film. The first reproducible, damage-free window—not any present numerical estimate—should determine the carrier and roller design.

### 7.2 Complete-cycle timing and multi-head bounds

Let $m$ be a dimensionless spacing multiplier. For provisional inputs of $v = 1$ m s$^{-1}$ and $t_g = 180$ s,[8] $v\, t_g$ = 180 m. For the illustrative geometric screen, I set the rounded base allocation $s_0 = 200$ m ≈ $1.11 \times v\, t_g$ and use $s = m\, s_0$. The corresponding geometric allocation count is $N_{geom} = \mathrm{floor}(L/s)$. At $m = 1$, this gives ten allocations on a 2 km track (Supporting Information, Figure S3). Values $m = 2$ and 3 give five and three allocations, respectively; the last case leaves 200 m unallocated. These are geometric upper bounds, not production units. Every wake must again nucleate, align, merge essentially seamlessly, and reach the declared product criteria before the next pass. No nonintersecting carrier-and-product path has been defined for simultaneous traveling heads. I therefore use one head as the sole illustrative calculation basis. This is a geometric calculation, not a production forecast; no delivered output is credited until the complete product path has been demonstrated.

### 7.3 Thickness and handling

Film thickness may help handling, but it is not a general solution. Interlayer slip makes few-layer bending stiffness dependent on curvature, whereas fracture remains controlled by flaws, support, and loading geometry.[41–43] Published bilayer, trilayer, graphite, and multilayer-hBN growth demonstrates accessible thicknesses, not high-speed hot pickup.[17, 44–46] Thick rBN may be easier to support yet harder to bend; a monolayer may bend readily yet be difficult to carry without support. I therefore retain monolayer graphene as the quantitative reference and treat thickness, support, and flaw population as variables to be measured.

## 8. Substrate fabrication at kilometer scale

A flat, smooth, 2 km × 30 cm M(111) substrate has, to the best of my knowledge, no direct manufacturing precedent. The closest platform is second-generation high-temperature-superconductor coated conductor: a long, flexible, biaxially textured template produced reel-to-reel.[47] I consider three routes: (i) deposit a candidate Ni(111), Ni-rich Ni–Cu(111), or other selected M(111) film on such a template and sharpen or seed its orientation; (ii) pass a metal web through a stationary seeded-recrystallization zone; or (iii) join many aligned segments (Figure 3). All require major advances. The target area is ~600 m², about five orders of magnitude larger than reported contact-free single-crystal metal foils (~32 cm²) and large-area liquid-Cu growth (~20 cm²).[48, 49] The required azimuthal spread is also near zero, not merely a few degrees.

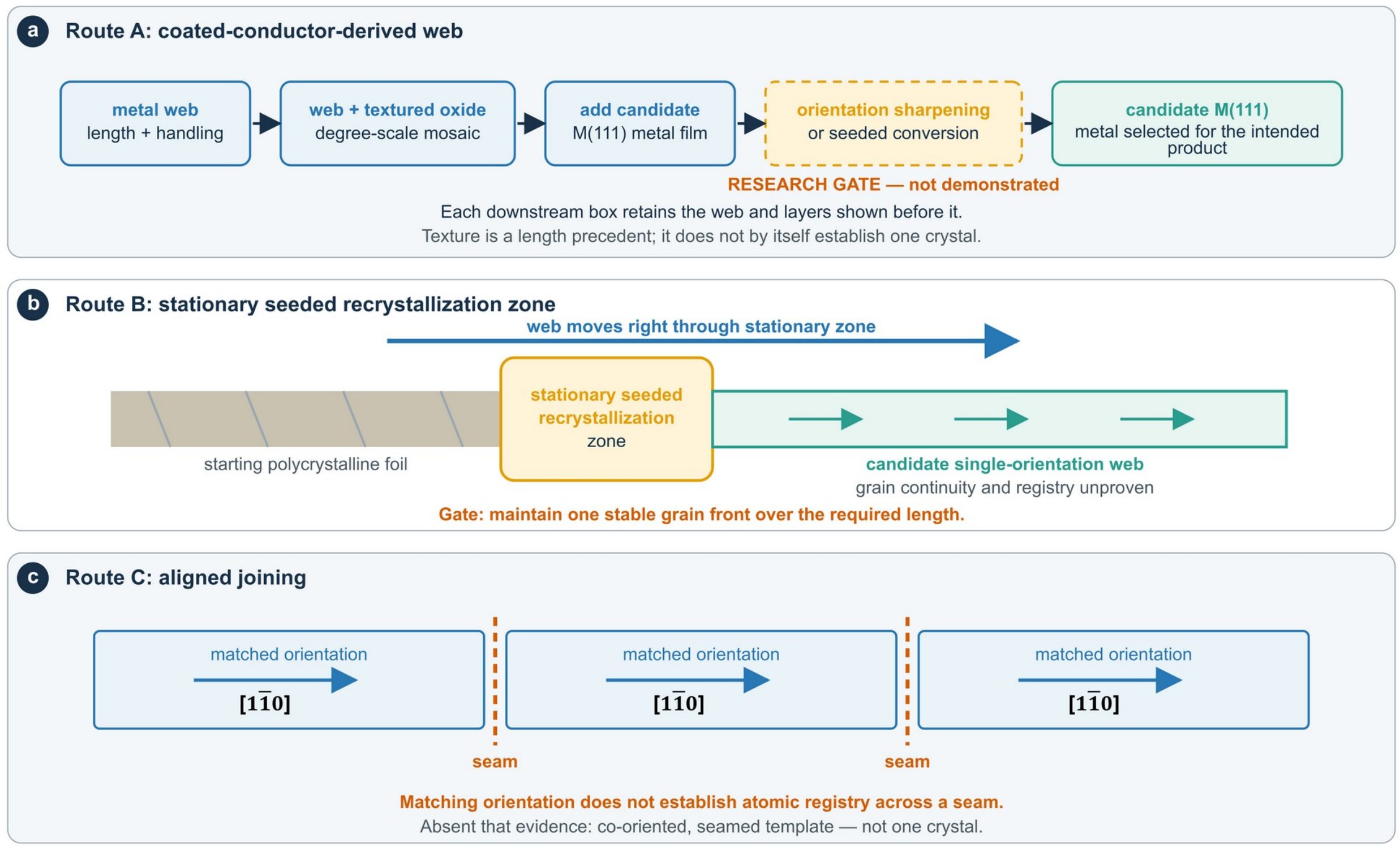

Figure 3. Three unproven routes toward a long M(111) surface. (a) A coated-conductor-derived web provides length and handling. Each downstream box retains the preceding web and layers: a biaxially textured oxide is added to the web, and a candidate M(111) metal film is deposited on that stack. The oxide does not by itself establish one crystal, and orientation sharpening or seeded conversion remains a research gate. The selected metal could be Ni(111), Ni-rich Ni–Cu(111), or another M(111) matched to the product. (b) The web moves through a stationary seeded-recrystallization zone. Success requires one grain and its registry to remain stable over the required length. (c) Matching orientation across joined segments does not establish atomic registry across a seam; absent that evidence, the product is a co-oriented, seamed template. The [1$\bar{1}$0] arrows show a common in-plane direction. The dashed orange box marks an unproven sharpening or conversion step; the dashed orange vertical lines mark seams. All routes are conceptual and not to scale.

## 8.1 Epitaxial Ni(111) films on a ceramic template

Coated conductors provide length, flatness, and reel-to-reel handling. A polished Hastelloy or steel tape carries barrier layers and a biaxially textured oxide, commonly IBAD-MgO, followed by an epitaxial functional layer. The cited work reached ~50 m $h^{-1}$ on ~1 cm-wide tape with a few-degree in-plane texture,[47] and industry has extended the platform to kilometer lengths. Two missing steps are decisive: a flexible oxide stack that templates a continuous (111) metal surface, and sharpening the in-plane texture from a few degrees to the level needed for seamless graphene or hBN merging. The c-plane-sapphire/Cu(111) result is a rigid-substrate precedent,[50] not evidence for either step on flexible tape.

A few-degree IBAD mosaic is textured, not single-crystalline, and would seed misoriented domains. A final seeded or surface-energy-driven recrystallization of the Ni film may sharpen the texture, but I know of no demonstration of such sharpening at the required length. C-plane sapphire can give better epitaxy but is slow to grow as wide ribbon.[51] It would also impose a one-time thermal-expansion mismatch with Ni. An MgO-based stack is better matched thermally, although its buffer layers and Ni thickness must still resist residual strain and delamination.

### 8.2 Single-crystal Ni(111) foil: conversion and seamless joining

The foil route starts with polycrystalline Ni and uses a seed to drive one (111) grain through the foil. Contact-free annealing has produced single-crystal Cu, Ni, Co, and Pt foils.[48] Related methods have produced ultraflat 6-inch Cu(111)[52] and seeded single-crystal Cu foils.[53] Recent room-temperature SCG growth used contact-free-annealed Cu(111),[5] confirming the substrate concept for graphene. The reported post-growth roughness ($R_q$ ≈ 1.8 nm) is consistent with degradation during hydrogen enrichment or another process step, but the study does not identify the cause. Reuse therefore requires cycle-resolved measurements of roughness, composition, and orientation.

A 2 km foil would require either continuous seeded recrystallization in a stationary zone through which the web moves or joining many aligned segments (Figure 3b,c). Only the first could, in principle, yield one translationally continuous crystal, and a stable kilometer-long grain front is unproven. In the joining route, EBSD-aligned segments could be diffusion-bonded, but matching orientation does not establish atomic registry across a seam. Partial precedents include Cu-to-sapphire bonding and seamless graphene merging on nearly commensurate Cu(111) domains,[10, 54] but they do not demonstrate end-to-end single-crystal foil joining. A joined substrate should therefore be called a "co-oriented, seamed template" unless atomic continuity is shown.

Here, "single crystal" is an operational, measurement-limited designation. For the metal, I require a stated azimuthal spread together with tests of translational continuity across the mapped area; EBSD, LEED, and X-ray diffraction widths alone do not exclude an incoherent seam. For the product, "near-single-crystal over one meter" is an aspirational designation for a full-width 1.00 m segment in which no disallowed boundary is detected at the stated resolution and sampling level. I do not assign a universal acceptable grain-boundary density. That value depends on the application. A film with grains hundreds of micrometers across may be valuable, but its measured grain-size distribution and boundary density should be reported rather than hidden inside the term "near-single-crystal." The separate target of "near-zero nonboundary macroscopic defects per meter" excludes grain boundaries from the defect count and reports tears, holes, voids, collapsed folds, and uncovered regions above a declared detection threshold. Continuous optical, scattering, or electrical inspection can screen macroscopic defects. Boundary-sensitive and, for hBN, polarity-sensitive measurements should calibrate those proxies by statistically declared sampling. The report should state the mapped area, spatial and angular resolution, sampling rule, and detection limit. A nondetection does not prove zero defects.

### 8.3 Liquid-mediated growth as a separate route

Liquid-mediated growth is pertinent, but two cases must be distinguished. Yang et al. reported thick single-crystal rBN by solid-liquid-interface-mediated epitaxy on Ni(111)/sapphire and proposed that a thin Ni–Si surface liquid stores B and N while the underlying solid Ni(111) preserves registry.[19] I consider this NUS archived work an important example of using a liquid in single-crystal BN growth. A free liquid-metal bath is different because it removes the solid crystallographic template. Liquid Cu is more credible for monolayer graphene; liquid Ni favors a larger carbon inventory and has produced bulk graphite during cooling.[23, 49, 55]

A refreshed free liquid surface may reduce persistent scratches, but bubbles, oxide or slag skins, evaporation, convection, capillary waves, and misregistered domain collisions create new defect channels. Direct hot lift-off and carrier capture must be demonstrated on a static bath before scale is considered. Self-aligned arrays are not necessarily single crystals.[56]

Many liquid-route parameters are unknown, and it is not possible or useful to solve them all in this Perspective. Section S7 and Table S7 instead order the first discriminating experiments. The analysis distinguishes the liquid-mediated rBN precedent of Reference 19 from a free-liquid bath and does not assume that either shares the solid-track apparatus.

Among solid routes, epitaxial Ni(111) on a textured web is more manufacturable at kilometer length, whereas converted foil offers higher local crystallinity. Neither yet meets both requirements. Several wafer-scale solid-film precedents sharpen the choice. Deng et al. repeatedly grew wrinkle-free SCG on a 500-nm single-crystal Cu(111) film on 4-inch sapphire and reused the same substrate through three growth-bubbling-transfer cycles.[57] They reported that the Cu roughness and graphene quality did not degrade. Cu90Ni10(111) was reported to enable 10 min growth of 4-inch graphene wafers, with 25 wafers processed in one cycle.[58] Full-coverage bilayer graphene was reported in 30 s on a 2-inch alloy film,[12] and composition-tunable single-crystal alloy (111) films with sub-0.2-nm roughness were reported in 2026.[59] The Deng study is direct evidence for batch reuse of a single-crystal metal film on sapphire. Its ~2 h full-coverage growth, cooling, PMMA support, and electrochemical bubbling transfer do not establish rapid regrowth, hot release, or rapid substrate regeneration. The faster preparation and growth precedents are References 12, 58, and 59. None of these studies establishes a kilometer-long translationally continuous crystal or hot delamination. I would begin the integrated program on a ~10-cm bulk or converted Ni(111) strip because that isolates the hot-release and regrowth physics without the textured-web mosaic. If S0–S2 pass, the scale ladder should be a 1-m converted Ni(111) strip, a 10-m hybrid web, a 200-m module, and only then a 2-km line. A long epitaxial Ni(111) web followed by seeded zone recrystallization remains the most plausible scale route, but it is a research hypothesis.

## 9. Atmosphere, contamination, and substrate conditioning

The tunnel atmosphere must keep M(111) metallic and sustain the selected growth chemistry. $H_2$-rich, Ar/$H_2$, and other validated mixtures should be compared; the required $H_2$ fraction has not been established for repeated peel-regrow operation. The background growth atmosphere is separate from the optional directed gas at the peel line. If a local gas is used, its shroud must limit exchange with the hydrocarbon growth zone, with composition sensors before and after the front. A conditioning zone should restore the specified growth atmosphere and use $H_2$ annealing or another treatment only if measurements show it is needed. The gas manifold and support system must also keep every full-width segment inside the temperature, pressure, reactant-partial-pressure, carbon-activity, flatness, and local-slope windows established on coupons. The first coupled transport and thermal-mechanical model should translate those measured windows into allowable manifold pressure drop, feed spacing, support spacing, sag, buckling, and axial and transverse temperature gradients. Exceeding any coupon-defined window is a gate failure, not a later optimization. Surface morphology, composition, orientation, regrowth delay, and film quality must be tracked cycle by cycle. Segmentation, automatic isolation of flammable gas when used, leak detection, purge capacity, and bounded hot-gas inventory are facility gates.

Bulk impurities such as S, P, and O can segregate to the hot surface, poison growth, and change adhesion. Their surface concentrations may decrease through depletion or increase through continued segregation. High-purity metal, in-line gettering or cleaning, and periodic regeneration must therefore be evaluated quantitatively. The dissolved-carbon inventory also changes as film is removed, so gas-phase carbon should remain the primary feed and the bulk reservoir should be monitored and recharged.

## 10. Throughput and scaling

Before yield is considered, one complete traverse exposes the usable substrate area $A = wL$ and can therefore produce at most that film area. For one head, define the mechanical cycle $T_{move} = L/v_p + L/v_r + t_{end}$ and the effective cycle $T_{eff} = \max(T_{move}, t_g)$. The gross rate is $Q_1 = A/T_{eff}$; delivered output must also include product yield and uptime. Figure 4 summarizes the complete 2 km cycle and three module-length screens. The timing expression does not establish product quality. The post-coverage interval is $T_{eff} - t_g$; for the illustrative 2 km case it is ~4020 s. The process passes only if repeated complete cycles yield accepted films at the scheduled peel. For $L$ = 2000 m, $w$ = 0.30 m, $v_p = v_r$ = 1 m s$^{-1}$, and $t_{end}$ = 200 s, the illustrative gross output is about $1.2 \times 10^4$ m$^2$ day$^{-1}$. This is an order-of-magnitude design screen, not a forecast.

**Complete-cycle timing and module-length screens**

One head; provisional 180 s full-coverage input [8]

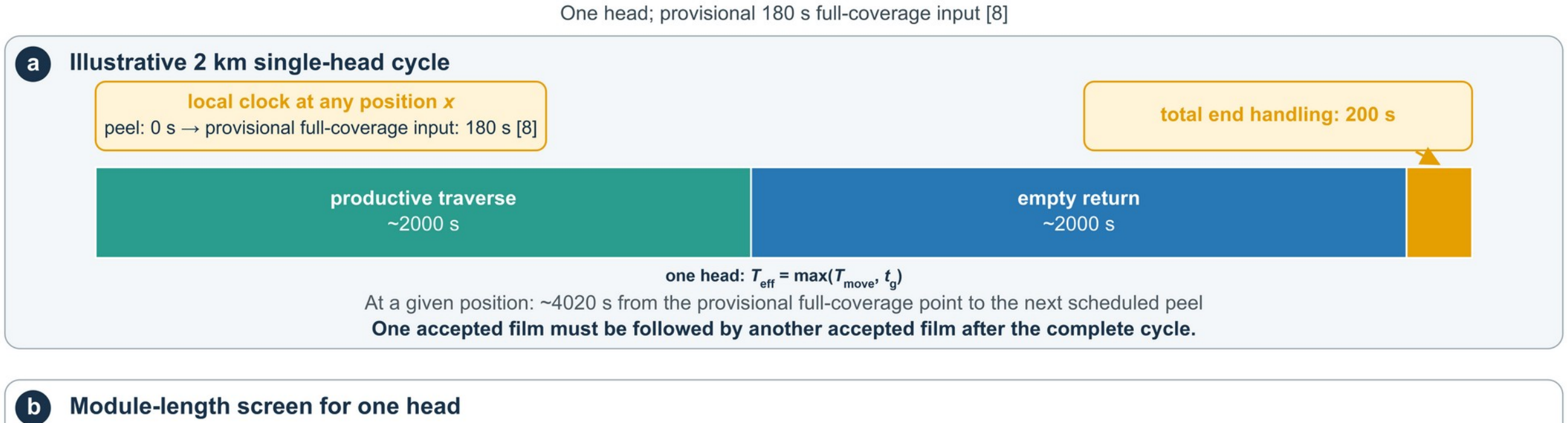

**b Module-length screen for one head**

$w$ = 0.30 m; productive and return speeds = 1 m s$^{-1}$; end handling = 200 s

| module length (m) | cycle time (s) | productive duty factor | gross geometric output (m$^2$ day$^{-1}$, before yield) | enclosure-wall energy screen (kWh per gross m$^2$) |
|---|---|---|---|---|
| **200** | 600 | 0.33 | **~9,000** | 1–5 |
| **500** | 1,200 | 0.42 | **~11,000** | 3–10 |
| **2,000** | 4,200 | 0.48 | **~12,000** | 10–40 |

Boundary: enclosure wall only (SI Table S6); excludes carrier, gas handling, cooling, yield, and downtime.

**All values are illustrative screens, not performance forecasts.**

Figure 4. Complete-cycle timing and module-length screens for one head at $w$ = 0.30 m, $v_p = v_r$ = 1 m s$^{-1}$, total end handling $t_{end}$ = 200 s, and provisional $t_g$ = 180 s.[8] Reference 8 identifies work in progress in our group; a public preprint or archive will replace the provisional citation when available. For these one-head rows, $T_{move} = L/v_p + L/v_r + t_{end}$, and $T_{eff} = \max(T_{move}, t_g) = T_{move}$. The productive duty factor is $(L/v_p)/T_{eff}$, and gross geometric output is $86{,}400 \times wL/T_{eff}$ in m$^2$ day$^{-1}$. (a) The illustrative 2 km cycle contains ~2000 s productive travel, ~2000 s empty return, and 200 s total end handling. For this screen, the provisional 180 s full-coverage input is conditionally assigned after each peel; the next scheduled peel is ~4200 s later, leaving ~4020 s after the conditional coverage point. (b) Shorter modules reduce cycle time and wall-energy exposure but lower the productive duty factor. Output values are gross geometric areas before yield. Enclosure-wall energy uses the provisional boundary in Table S6 and excludes carrier manufacture, gas handling, cooling, yield, downtime, and other balance-of-system terms. All values are design screens, not forecasts.

The hot enclosure is likely to dominate energy use. At the illustrative single-head point, interfacial separation is only ~0.3 W, and the representative carrier requires roughly 40 kW during the productive traverse. For a 2 km enclosure with a 4.5 m exposed perimeter, the wall area is ~9000 $m^2$. A provisional wall-loss range gives roughly 5–20 MW, or about 10–40 kWh per gross square meter. If four serial yields are each assumed to be 90% and overall equipment effectiveness is assumed to be 50%, the conforming fraction is $0.9^4 \times 0.5 \approx 0.328$ and the wall-only result is roughly 30–120 kWh per conforming square meter. These assumptions are illustrative. They are not performance forecasts, competitiveness claims, or a life-cycle assessment, and they should be replaced by experimental and pilot-line data. Heating, pumping, exhaust treatment, cooling, carrier manufacture, and downtime remain incompletely bounded. Table S6 states the present information boundary.

Delivered output must be reported as conforming product, not geometric area. The accounting boundary includes coverage, crystallinity, layer-number and peel yields; carrier capture and downstream release; uptime; wall losses; gas heating and pumping; cooling; and carrier production or recycling. Because several terms are unmeasured, the present calculation identifies dominant measurements rather than a life-cycle energy result.

Length is a design variable, and the 2 km case is not presently preferred. At the stated speeds and 200 s end handling, 200, 500, and 2000 m modules give cycles of 600, 1200, and 4200 s; illustrative gross outputs of about 9,000, 11,000, and 12,000 $m^2$ $day^{-1}$; and wall-energy screens of roughly 1–5, 3–10, and 10–40 kWh per gross square meter. Shorter modules reduce the post-coverage interval, carrier inventory, thermal expansion, wall loss per square meter, and common-mode failure exposure, but fixed end handling reduces their productive duty factor. Parallel modules a few hundred meters long may therefore outperform one 2 km line. Experiments, uptime, carrier logistics, and failure statistics must refine every estimate and decide the length after the local physics passes.

## 11. Decisive risks and required studies

The solid route should proceed through the ordered gates in Table 1. The table is not complete and is not intended to freeze the development sequence. It records the problems that can be identified now; a failure mode revealed by experiment, whether listed or not, becomes a new gate. The first two gates test hot release and repeated full-cycle growth on a mapped coupon-scale M(111) crystal. The third tests substrate integrity and continuity as scale increases; the later gates address chemistry, product handling, and scale. The near-single-crystal claim requires a stated mapped area, resolution, and sampling rule. I do not prescribe a universal acceptable grain-boundary density. If boundaries are detected, their density and the grain-size distribution should be reported. A separate per-meter count covers nonboundary macroscopic defects above the declared detection threshold.

| Gate | Question | Measurement / pass criterion | Stop or redirect |
| --- | --- | --- | --- |
| S0 — Hot release | Can a stable peel front traverse useful width and speed? | Map effective G, front variance, tears, residue, and film damage versus v, T, atmosphere, and width. Compare no directed crack-tip gas first, then Ar, then a reactive gas only if useful. | Stop the repeated hot-release route if no reproducible damage-free window exists. |

| Gate | Question | Measurement / pass criterion | Stop or redirect |
|---|---|---|---|
| S1 — Repeated growth | Does each complete cycle yield another accepted film? | Repeat growth, dwell, peel, and regrowth. At each scheduled peel, report coverage, layer number, mapped boundaries, grain-size distribution if boundaries are present, and nonboundary macroscopic defects above the declared detection threshold. Test feed gating only if the full cycle fails. | Stop scale-up if product quality or catalyst state degrades through repeated cycles and no workable change in chemistry, gating, or module length restores it. |
| S2 — Substrate integrity and continuity | Is M(111) one crystal at the stated detection limit, and does its state remain acceptable through cycling? | Map azimuthal spread and translational registry before and after repeated cycles. For roll acceptance, inspect every consecutive full-width 1.00 m segment and each segment junction at declared spatial and angular resolution; track roughness and composition cycle by cycle. | A detected disallowed boundary prevents the near-single-crystal designation. Degradation of crystal state, roughness, or composition stops scale-up unless corrected. A nondetection sets only an upper bound. |
| S3 — hBN-specific | Can phase, layer number, and sublattice be controlled? | Select the hBN product first, then map M(111), B/N activity, borides, layer or thickness, polarity, inversion or antiphase boundaries, hot release, repeated growth, apparatus, and scale strategy; see Table S5. | Failure excludes hBN from this platform, not necessarily graphene. |
| S4 — Thermal, flow, and chemistry stationarity | Does every segment remain inside the measured process and flatness windows? | Close S, C, H, O, B, N, and alloy inventories; map axial and transverse temperature, pressure, gas composition, carbon activity, roughness, sag, and local slope. Separate the growth atmosphere from optional local peel gas. | A segment outside a coupon-defined chemistry, temperature, pressure, or flatness window stops long-run development. |
| S5 — Carrier path | Can the product be captured, cooled, and released? | Demonstrate capture, supported cooling, collection, downstream release, contamination control, and reuse. Map carrier thickness, modulus, creep, CTE mismatch, roller radius, line tension, adhesion hierarchy, and release energy. | Do not credit delivered output until this route is demonstrated; multi-head output needs a separate nonintersecting path. |
| S6 — Scale | Does a shorter validated module justify greater length? | Use the ordered ladder: ~10 cm M(111), 1 m converted foil, 10 m hybrid web, then compare 200, 500, and 2000 m modules by repeated-cycle product quality, rounded output and energy screens, uptime, carrier logistics, and failure exposure. | Do not advance a length step until S0–S5 pass at the preceding scale. |

Table 1. Ordered go/no-go gates for the solid-substrate route. S0–S4 are scientific kill gates; S5 must close the single-head product path. Track length, roller radius, carrier thickness, and geometric head count are optimizable only after these requirements are met.

This ordering separates kill criteria from engineering variables. Failure of S0, S1, or S2 removes the physical basis for repeated single-crystal graphene production; failure of S3 removes hBN from the defined extension; failure of S4 precludes stable long runs. Carrier choice is not a minor optimization: S5 requires a material that survives the thermal and mechanical window, adheres strongly enough for capture, releases cleanly downstream, and can be reused without contaminating the product. Roller radius and module length may then be optimized within the surviving process window. Multi-head output remains outside the credited baseline until its product paths close.

## 12. Conclusion

I asked whether SCG—and, as a separate case, SC-hBN—could be grown, hot-delaminated, and regrown repeatedly on a long, reusable single-crystal metal surface. Reference 8 identifies work in progress in our group that has yielded a 180 s full-coverage result. I use that result provisionally, and a public preprint or archive will replace the citation when available. The 30 s bilayer result of Reference 12 is a faster but different wafer-scale precedent. The decisive kinetic test is a repeated complete cycle: an accepted film must be present at the scheduled peel and an accepted film must form again afterward. Rapid hot release at 1 m $s^{-1}$ remains unvalidated. Graphene/Ni(111) has modest intrinsic adsorption energy, yet effective adhesion, optional gas-assisted decoupling, carrier mechanics, and tear statistics under process conditions are unknown.

The concept also lacks direct precedent for a meter-to-kilometer substrate that meets an operational single-crystal criterion. hBN offers real choices: Cu–Ni(111) may favor monolayer products, whereas Ni(111) can support thicker single-crystal films. It must pass separate phase, sublattice, layer, and hot-release gates, and it may require a different apparatus and scaling strategy. The liquid-mediated rBN result of Reference 19 is important because a thin liquid layer participated while solid Ni(111) supplied registry. A free-liquid bath remains a different, separately gated architecture.

The priority is therefore clear. First, using a mapped coupon-scale M(111) surface, establish a stable hot-release window across increasing speed and width. Second, on the same surface, show accepted films through repeated complete growth-dwell-peel-regrowth cycles. These two experiments determine whether the concept survives. If either fails, diagnose the cause and test local feed gating, a changed atmosphere, a different carrier or interface condition, or a shorter module. The substrate, atmosphere, carrier, quality-control, energy, and scaling analyses elsewhere in this Perspective address other requirements that could become decisive after those tests. They are necessarily incomplete; their value is to make present assumptions explicit and identify measurements, not to claim a perfected process design. If experiment reveals an unanticipated coupling, the design and gate sequence should change. Third, if the two central experiments pass, extend and requalify the reusable M(111) surface through the proposed scale ladder while closing the surface-chemistry and carrier paths. Only then should module length be selected. hBN should follow its own M(111), thickness, release, apparatus, and scale choices. I hope that these experiments are taken up, including by others, whether or not the architecture as a whole survives them.

## Acknowledgments

This work was supported by the Institute for Basic Science (IBS-R019-D1), Republic of Korea. High-speed, high-temperature peeling studies motivated by this concept are being considered with Prof. Seunghwa Ryu (KAIST; molecular-dynamics simulation and machine-learning upscaling of interfacial fracture), Prof. Nicola M. Pugno (University of Trento; multiple-peeling and rate- and temperature-dependent fracture theory), and Prof. Yarjan Samad (Khalifa University). I thank John Knapman and Adrian Nixon for comments, and Yarjan Samad for his detailed reading and comments. I sincerely thank Alisher Sultangaziyev for assistance in preparing the figures.

During preparation and revision of this work, the author used Claude (Anthropic), accessed through the Claude Cowork desktop application (2026), and Codex (OpenAI), accessed through the Codex desktop application (2026), to assist with literature searching and organization, drafting and editing, reference checking against source documents, adversarial review for internal consistency, and development of initial conceptual drafts of the schematic figures. Alisher Sultangaziyev manually reconstructed, scientifically corrected, and finalized the figures using Microsoft PowerPoint and Adobe Photoshop under the author's direction. The figures are conceptual schematics and do not contain or alter experimental data. The author reviewed and edited all outputs retained in the manuscript, made all scientific judgments, and takes full responsibility for the article.

## Conflict of Interest

The author declares no conflict of interest.

## Author Contributions

Rodney S. Ruoff conceived the manufacturing concept, identified hot film removal as the likely rate-limiting step in the proposed architecture, recognized usable single-crystal substrate area as the principal geometric determinant of material produced per growth cycle, framed the quantitative analysis, and wrote the manuscript.

## Data Availability Statement

No new experimental data were generated for this Perspective. Published quantitative values and source materials discussed in the article are available in the cited references. The 180 s full-coverage value is drawn from work in progress in our group and is used here only as a provisional design input; Reference 8 lists the contributors, and a public preprint or archive will replace the provisional citation when available.

## Supporting Information

Supporting Information is available from the Wiley Online Library or from the author.

## Table of Contents Entry

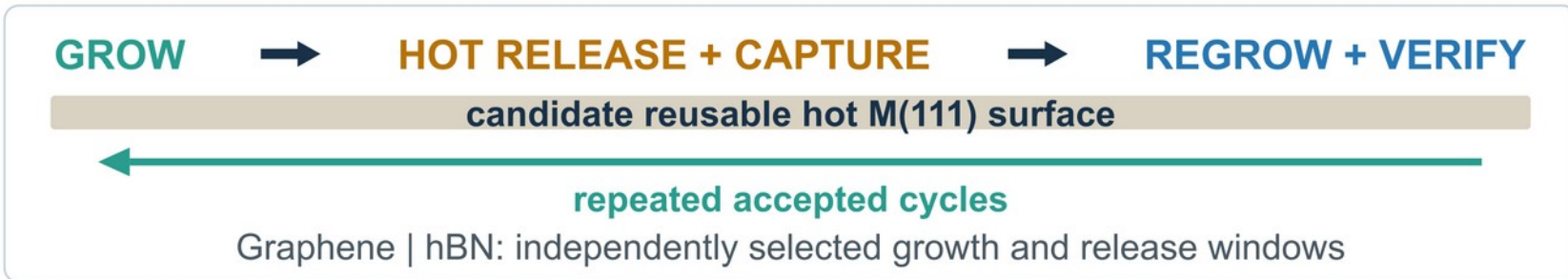

This Perspective examines repeated growth and hot delamination of single-crystal graphene on a reusable M(111) surface, with hBN treated separately. Once rapid growth is available, hot removal is likely to set the cycle rate, whereas usable single-crystal growth area sets the material produced per cycle. Repeated accepted growth-peel-regrowth cycles determine whether the proposed architecture survives.

# Supporting Information

## Repeated Growth and Hot Delamination of Single-Crystal Graphene: A Two-Kilometer Process-Design Assessment with hBN as a Separate Case

*Rodney S. Ruoff**

R. S. Ruoff, Center for Multidimensional Carbon Materials (CMCM), Institute for Basic Science (IBS), Ulsan 44919, Republic of Korea; Department of Chemistry, Department of Materials Science and Engineering, and School of Energy and Chemical Engineering, UNIST, Ulsan 44919, Republic of Korea. E-mail: ruoff@unist.ac.kr

*Citation key: [n] denotes a main-text reference; [Sn] denotes a Supporting Information reference listed at the end of this document.*

### S1. Adsorption state and separation energy at the graphene-Ni(111) interface

Section 5 of the main text distinguishes adsorption energy, bond character, and effective peel energy. RPA/ACFDT calculations for graphene/Ni(111) give π–d($z^2$) hybridization at 2.17 Å and an adsorption energy of 67 meV per C (~0.41 J m−2).[26] The binding energy includes adsorption-induced exchange changes and van der Waals correlation.[26] The same calculation gives a second, physisorptive minimum near 3.3 Å at ~60 meV per C.[26, 27] Independent RPA work gives a near minimum at ~2.3 Å and a far minimum at ~3.25 Å, with an energy difference of ~5–8 meV per C after k-point convergence.[27, S1] These results establish that a short equilibrium separation and orbital hybridization can coexist with a modest adsorption energy. They do not give the mechanical energy required by a moving peel front.

The quantitative interpretation is method-dependent. RPA may underbind covalent interactions,[27] whereas Wannier analyses describe the near state as chemical in character despite its modest binding energy.[28] LZK dispersion models reproduce the far ~3.3 Å physisorption well.[S2, S3] XDM calculations give both minima and predict that thermal expansion of Ni stabilizes the near state.[S4] A DFT-D study instead favors chemisorption by several kJ $mol^{-1}$ per C.[S5] Molecular-beam-scattering calculations also indicate strong graphene-Ni coupling,[S6] and the adsorption state changes with facet.[S7] I therefore use a mixed weak-chemisorption/physisorption description and keep adsorption state separate from effective adhesion. None of these calculations includes a hot, advancing peel front tested with no directed gas, Ar, or a reactive gas.

Table S1. Computed adsorption energies and equilibrium separations for graphene on Ni(111), together with one measured binding distance. Energies converted at 1 meV/C = 0.00611 J $m^{-2}$ (graphene areal density 3.82 × $10^{19}$ C $m^{-2}$). LDA (overbinds) and pure vdW-DF (underbinds) are shown for context.

| Method | d (Å) | E_ads (meV/C) | ≈ J/m² | Source |
|---|---|---|---|---|
| RPA (ACFDT), near minimum | 2.17 | 67 | 0.41 | Mittendorfer 2011 [26] |
| RPA, far (physisorption) minimum | 3.3 | 60 | 0.37 | Mittendorfer 2011 [26] |
| RPA (Thygesen group) | 2.19 | 70 | 0.43 | Olsen 2011;[27] [S1] |
| RPA, far minimum | 3.25 | ≈5–8 meV/C weaker than near minimum | — | Olsen [27, S1] |
| PBE (no vdW) | 2.17 | +9 to +15 (unbound) | — | Mittendorfer |

| Method | d (Å) | E_ads (meV/C) | ≈ J/m² | Source |
|---|---|---|---|---|
| | | | | 2011 [26] |
| LDA (artifact) | 2.00 | ~196–200 | ~1.20 | Silvestrelli 2015 [28] |
| vdW-DF | 3.5–3.7 | 37–44 | 0.23–0.27 | Olsen;[27] Silvestrelli [28] |
| optB88-vdW | 2.18 | 71 | 0.43 | Silvestrelli 2015 [28] |
| optB86b-vdW | 2.12 | 112 | 0.68 | Silvestrelli 2015 [28] |
| DFT/vdW-WF2s1 (Wannier) | 2.12 | 129 | 0.79 | Silvestrelli 2015 [28] |
| PBE+vdW (LZK model), far well | ~3.3 | MAE 7 vs RPA | — | Tang 2019 [S3] |
| Experiment (distance only) | 2.11 ± 0.07 | — | — | Gamo 1997 [S8] |

Reference values are ~48 meV per C (~0.29 J m−2) for RPA graphite interlayer cohesion and 62 meV per C (~0.38 J m−2) for RPA graphene/Cu(111) at 3.25 Å.[27] The calculated ~0.41 J m−2 value for graphene/Ni(111) is only modestly larger. This comparison helps set an intrinsic energy scale; it does not predict the effective peel energy.

## S2. Measured effective adhesion

The intrinsic thermodynamic work of separation (~0.4 J m−2 by RPA) is a lower bound on mechanically measured effective adhesion. Peel and fracture measurements include rate-dependent dissipation, substrate plasticity, roughness, pinning, mode mixity, and contamination. Reported values span ~0.7–6.8 J m−2 (Table S2). One StereoDIC blister study on Ni reported ~6.8 J m−2 and interpreted the value as near-covalent;[29] it has not been independently corroborated. Roll-to-roll estimation gave 1.2–2.6 J m−2,[S9] and high-rate delamination reached ~6 J m−2.[33] Nanoscratch measurements gave still larger apparent values,[S10] but later work attributed much of that response to metal plasticity rather than interfacial separation.[S11] Roll-to-roll dry peeling demonstrates web-scale transfer,[S12] not the high-temperature adhesion or speed required here.

***Table S2.*** Representative measured effective adhesion energies for graphene on metals (method-dependent).

| System / method | Effective adhesion (J/m²) | Source |
|---|---|---|
| Graphene/Cu, as-grown (DCB) | 0.72 | Yoon 2012 [S13] |
| Graphene/Cu, blister test | 0.7–1.5 | Xin 2017 [S11] |
| Graphene/Ni, StereoDIC blister | ~6.8 (near-covalent) | Chang 2019 [29] |
| Graphene, roll-to-roll (in-process) | 1.2–2.6 | Hong 2023 [S9] |
| Graphene/Cu, rate-dependent delamination | ~6 (high speed) | Na 2015 [33] |
| Graphene, nanoscratch (apparent) | 12.8 (Cu); 72.7 (Ni) | Das 2013 [S10] |
| Graphene/Cu, large-area transferred blister | ~0.3–0.5 (transferred; < as-grown) | Cao 2014 [S14] |

## S3. Substrate fabrication at kilometer scale (extended)

Section 8 treats the substrate as a separate development program. The target is a flat, smooth M(111) surface, 30 cm wide and ~2 km long. Route A uses a reel-to-reel textured ceramic template derived from second-generation high-temperature-superconductor coated conductors, then deposits a candidate Ni(111), Ni-rich Ni–Cu(111), or other selected M(111) metal film.[47] This platform provides length and web handling, but metal-film formation is a separate step, and its few-degree in-plane mosaic must be sharpened to the level required for seamless graphene or hBN merging. Route B converts Ni foil through contact-free grain growth and extends the converted orientation by passing the web through a stationary seeded-recrystallization zone.[48] Route C uses aligned joining. Joined pieces remain a co-oriented, seamed template unless translational continuity is demonstrated across every joint. Cu(111) bonding and graphene stitching on nearly commensurate Cu(111) domains are partial precedents only.[10, 54] Deng et al. provide a rigid-wafer reuse precedent: a 500 nm twin-free single-crystal Cu(111) film on 4 inch oxygen-pretreated sapphire was used for three graphene growth-bubbling-transfer cycles.[57] The Cu roughness and graphene quality were reported not to degrade, but full graphene coverage required ~2 h and transfer used PMMA and electrochemical bubbling after cooling. This establishes batch reuse, not rapid regrowth, hot release, or rapid substrate regeneration. The hot support and manifold must keep every segment inside coupon-defined windows for flatness, local slope, temperature, pressure, reactant partial pressure, and carbon activity. The decisive measurements are orientation, translational registry, roughness, creep or warp, thermal gradients, flow uniformity, and stability during long high-temperature operation.

***Table S3.*** Comparison of the two substrate-fabrication routes. Bracketed numbers are main-text references.

| Attribute | Route A: epitaxial Ni(111) film on ceramic template | Route B: single-crystal Ni(111) foil |
|---|---|---|
| Length capability | km, reel-to-reel (proven for 2G-HTS) [47] | dm–m today; km needs continuous zone or tiling [48, 52] |
| Width | 30 cm target; continuous width at kilometer length is unproven | 30 cm target; continuous width at kilometer length is unproven |
| Crystal quality | Biaxially textured (~3° mosaic), not single-crystal; must sharpen | true single crystal locally; seams risk grain boundaries [10] |
| (111) out-of-plane | textured (111) template + surface-energy bias [50] | natural (lowest-energy fcc facet) + seed [53] |
| Flatness | Web-compatible; final Ni(111) roughness must be measured | Local flatness demonstrated; long-run flatness is unknown |
| Robustness at 1050 °C | High-temperature platform precedent; Ni/oxide stack, support flatness, sag/buckling, and flow/thermal uniformity at 1050 °C are unproven | Foil must resist creep, warp, support imprinting, and axial/transverse thermal gradients |
| Maturity | Manufacturing platform exists, but not at required crystal quality; test after local physics passes. | First demonstrator: ~10 cm bulk or converted Ni(111); then 1 m converted foil. |
| Key risk | Sharpen mosaic and preserve the oxide/metal stack through 10 m → 200 m → 2 km scale steps. | Maintain one advancing grain and hot-release/regrowth performance through the 10 cm → 1 m steps. |

I suggest one hybrid worth testing: a long epitaxial Ni(111) film on a textured web followed by seeded recrystallization. It could combine Route A's handling with Route B's local crystal quality. This is a research hypothesis; neither true single-crystal conversion nor stability at 2 km has been demonstrated.

## S4. Pickup, carrier, and quantitative assumptions

Section 7 of the main text defines the functions required for pickup and rollup rather than a final machine. This section preserves one candidate test layout and the calculations needed to size the first experiments. They are design screens, not performance predictions or a complete facility specification. Before hot peeling and rapid repeated cycling are attempted, it is not possible to enumerate every coupled constraint or failure mode. Any new failure exposed experimentally should revise the design and the gate sequence.

### S4.1 Candidate peel and capture arrangement

Figure S1 defines a continuously supported coupon-scale hot-peel experiment and the atmosphere sequence. The film remains supported by M(111) or by the laminated carrier; the carrier follows a single-pass path, not a loop. The test head remains stationary while the heated M(111) coupon translates to the left. A contact roller first presses the preheated carrier onto the still-supported film. The carrier-film composite then moves left from the crack front into a thermally isolated force-measurement path. The measured system force includes peel resistance, web tension, and roller losses, so each condition requires a no-peel or otherwise differential calibration. The first condition has no directed gas. If needed, Ar and then $H_2$ or activated H can be directed into the opened wedge. Force, front stability, film damage and residue, subsequent regrowth, and product acceptance are measured together.

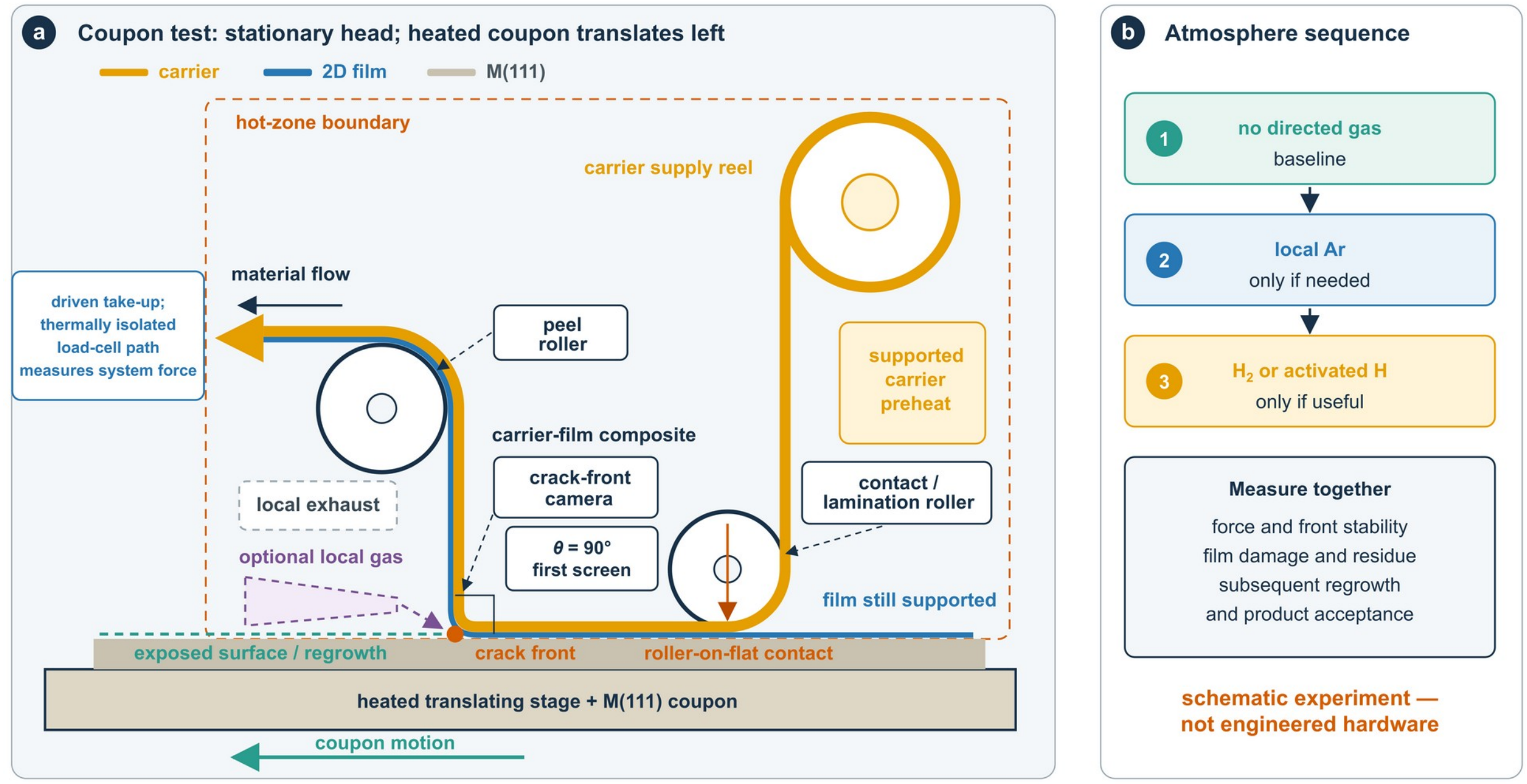

Figure S1. Schematic of a continuously supported coupon-scale hot-peel experiment; not to scale. Orange denotes the carrier, blue the 2D film, and beige M(111). The film remains supported by M(111) or by the laminated carrier, and the carrier follows a single-pass path. (a) The heated coupon moves left beneath a stationary head. A contact roller laminates the preheated carrier while the film remains supported; the carrier-film composite then peels at θ = 90° for the first force screen and enters the left-hand driven take-up through a thermally isolated load-cell path. Material-flow, coupon-motion, lamination-load, and gas arrows are separately identified. A camera records the crack front. The measured system force requires calibration for web tension and roller losses. (b) The first condition has no directed gas. If needed, Ar and then $H_2$ or activated H are tested. Every condition is judged by force and front stability, film damage and residue, subsequent regrowth, and product acceptance. The drawing defines an experiment, not a machine.

### S4.2 Peel geometry, speed, and force screen

Film bending is not expected to set the roller radius. The strain $\varepsilon \approx t/(2R)$ is ~$2 \times 10^{-7}$ for a five-layer film ($t \approx 1.7$ nm) on a 5 mm radius. Handling stability and peel-front sharpness therefore determine $R$; 10–50 mm is a starting range to test, not a design selection.

At 90°, the ideal inextensible-film relation gives $F/w \approx G$. Effective adhesion values of 1 and 6 J m$^{-2}$ correspond to ~1 and ~6 N m$^{-1}$, or ~0.3–1.8 N across 0.30 m. The load cell does not measure interfacial separation alone: web tension, roller and bearing losses, carrier bending, and thermal drift contribute. Each condition therefore requires a no-peel baseline or another differential calibration. Only the film near the crack line and the short span to the nip carry the peel load; the full 2 km sheet is never tensioned at once. Centimeter-scale SCG reached ~27.4 GPa (~9.2 N m$^{-1}$), whereas an edge-flaw projection gives ~13.7–18.4 GPa (~4.6–6.2 N m$^{-1}$).[2] These monolayer values do not establish the strength of the multilayer graphene or hBN considered here. At the upper adhesion screen, the projected breaking and peel loads overlap. Local pins, step bunches, particles, front skew, and carrier nonuniformity can further concentrate stress; the values therefore define measurements, not a safe operating window.

The assumed 1 m $s^{-1}$ speed is aspirational. Reported graphene and hBN transfer occurs at much lower speeds, including centimeter-per-second roll-to-roll peeling,[S12] sub-millimeter-per-second rate-dependent delamination,[33] and quasi-static blister tests.[29] Tape peeling reaches the meter-per-second range but can exhibit stick-slip between ~0.25 and 2.45 m $s^{-1}$.[34–36] Qualification should therefore use a width-speed-temperature-atmosphere ladder with synchronized force and front imaging.

**S4.3 Carrier heating, product handling, and defect statistics**

The interfacial separation power, $Gwv$, is ~0.3 W for $G$ = 1 J $m^{-2}$, $w$ = 0.30 m, and $v$ = 1 m $s^{-1}$. Facility heating and mechanical losses should therefore dominate the energy needed to separate the interface. For a representative 25 µm metal-like carrier with areal mass ~0.20 kg $m^{-2}$ and a roughly 1000–1200 K temperature span, a 30 cm web moving at 1 m $s^{-1}$ requires roughly 40 kW of sensible heating or cooling during the productive traverse. One 600 $m^2$ roll contains ~120 kg of this assumed carrier. With a 4200 s cycle, daily carrier throughput is ~2.5 t and the cycle-averaged sensible-heat rate is ~19 kW. These values do not select a material.

The area also makes rare defects important. One 2 km × 30 cm pass is 600 $m^2$. If every fatal pin were independent and spatially homogeneous, 95% zero-pin yield would require a density below $-\ln(0.95)/600 \approx 8.5 \times 10^{-5}$ $m^{-2}$. Real defects may cluster, and detection probability is below unity. This Poisson result is therefore an idealized screen; segmentation, crack arrest, in-line mapping, and measured spatial statistics belong in the experimental program.

A bare monolayer should not be assumed to survive an unsupported hot span. Figure S2 therefore uses the laboratory frame of the baseline architecture: M(111) is fixed and the carriage moves to the right. A carrier from a carriage-mounted supply reel is preheated while supported and pressed onto the still-supported 2D film ahead of the crack. The resulting carrier-film composite peels together, wraps tangentially around a peel roller, and travels rearward over a conceptual cooled support deck or roller train before force and tension control, edge guidance, and tangent entry to the take-up. The support geometry has not been selected, but the path does not require a long unsupported composite span. Thus, the film is not first peeled bare and then captured. Once laminated, the peel arm is a hot composite; carrier bending, creep, plastic work, roller friction, and differential contraction may exceed the interfacial contribution. Initial screening should bracket 10–25 µm Mo and thin high-temperature Ni-alloy foil, but no carrier, roller diameter, contact pressure, cooling length, detailed support design, take-up torque, or changing-roll-mass limit has been selected. The path ends with carrier-supported take-up; downstream release is a separate experiment.

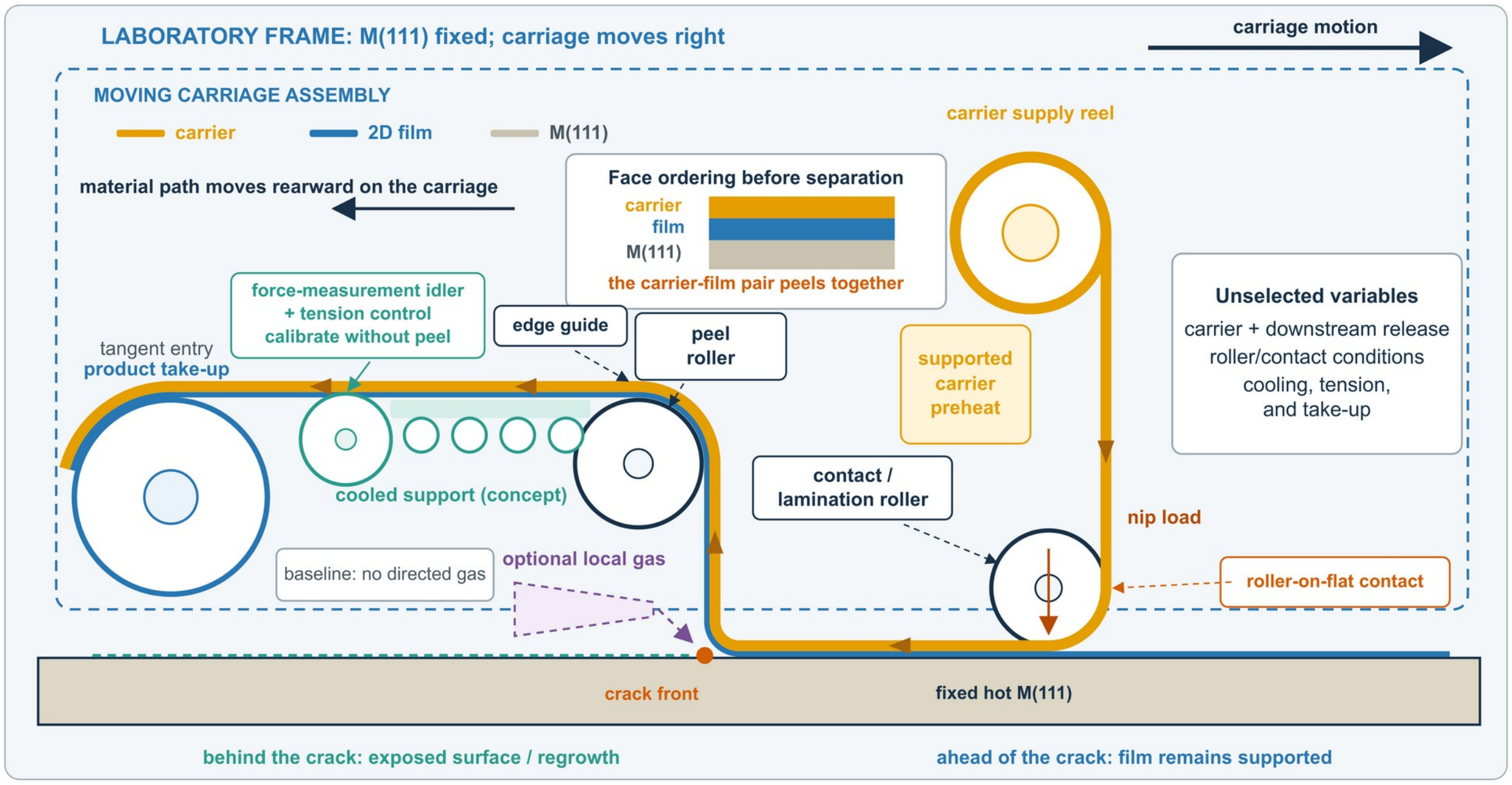

Figure S2. Schematic of a carrier-supported path through take-up on the moving carriage; not to scale. Orange denotes the carrier, blue the 2D film, and beige M(111). The dashed region is the boundary of carriage-mounted components, not the hot enclosure; fixed hot M(111) lies below it. In the laboratory frame the carriage moves right, whereas the carrier-film product moves rearward, to the left, on the carriage. The carrier is preheated while supported and pressed onto the still-supported film ahead of the crack. The carrier-film composite then peels together, follows tangent roller paths, and crosses a conceptual cooled support deck or roller train before tension control and tangent entry to the take-up. The support geometry remains unselected, but the drawing does not rely on a long unsupported span. The face order before separation is carrier/2D film/M(111). Optional local gas is directed into the opened wedge only if tested; the first condition has no directed gas. Carrier material, adhesion hierarchy, roller radii and pressure, cooling length, support design, take-up torque, changing roll mass, tension control, and downstream release remain unselected. The path ends with carrier-supported take-up; it is not a design for downstream release.

For areal mass 0.20 kg m$^{-2}$, heat capacity 0.5 kJ kg$^{-1}$ K$^{-1}$, temperature rise 1100 K, and speed 1 m s$^{-1}$, the supported heating length is ~0.55–1.1 m if net heat flux is 100–200 kW m$^{-2}$. This follows from $L_h = v\, m_A c_p \Delta T / q''$, and every input requires measurement. A useful carrier must give carrier/film adhesion above the effective film/metal separation energy during capture, release downstream without residue, and avoid damaging contraction strain after allowing for creep or slip.

Hot release may reduce cooling-induced wrinkles, but hot tensioning is not a general cure. Flattening a gentle wrinkle of amplitude $A$ and wavelength $\lambda$ requires strain of order $\varepsilon \approx \pi^2(A/\lambda)^2$. Sharp folds, thick rBN, and flaws require separate treatment. Film thickness may help handling, yet interlayer slip makes few-layer bending stiffness curvature-dependent, and fracture remains controlled by flaws, support, and loading geometry.[41–43] Published bilayer, trilayer, graphite, and multilayer-hBN growth demonstrates accessible thicknesses, not high-speed hot pickup.[17, 44–46]

**S4.4 Atmosphere and optional intercalation**

Intercalation may not be necessary. The first hot-peel measurement should use no directed gas at the crack line, followed by Ar and then $H_2$ or activated H only as a reactive comparison. If an added gas must act over $\delta$ = 10 mm ahead of a front moving at $v$ = 1 m s$^{-1}$, the available time is $\delta/v$ = 10 ms. A one-dimensional diffusion screen then gives an effective transport target of at least $\delta^2/(2t) \approx 5 \times 10^{-3}$ m² s$^{-1}$; any required adsorption or reaction must also occur within 10 ms. This screen does not constrain a successful no-directed-gas peel. Compare the conditions by effective adhesion, front stability, residue, etching, film damage, and regrowth delay. Failure of $H_2$ intercalation does not reject the process if no-flow or Ar-assisted peeling succeeds.

**S4.5 Quantitative assumptions and claim-evidence status**

Define the mechanical cycle as $T_{move} = L/v_p + L/v_r + t_{end}$ and the effective cycle as $T_{eff} = \max(T_{move}, t_g)$. The post-coverage interval is $T_{eff} - t_g$. Figure S3 adds an explicit geometric allowance: $s_0$ = 200 m ≈ 1.11 × $v\, t_g$ and $s = m\, s_0$, where m is a dimensionless spacing multiplier. This allocation screen does not establish simultaneous production heads. I do not impose a separate stability threshold if repeated full cycles yield accepted products at the scheduled peel and after regrowth. Feed-on/feed-off measurements are diagnostic if quality changes or if local gating is being considered. The calculations above and in Tables S4–S6 use conditional inputs; they identify measurements and do not constitute a complete facility design.

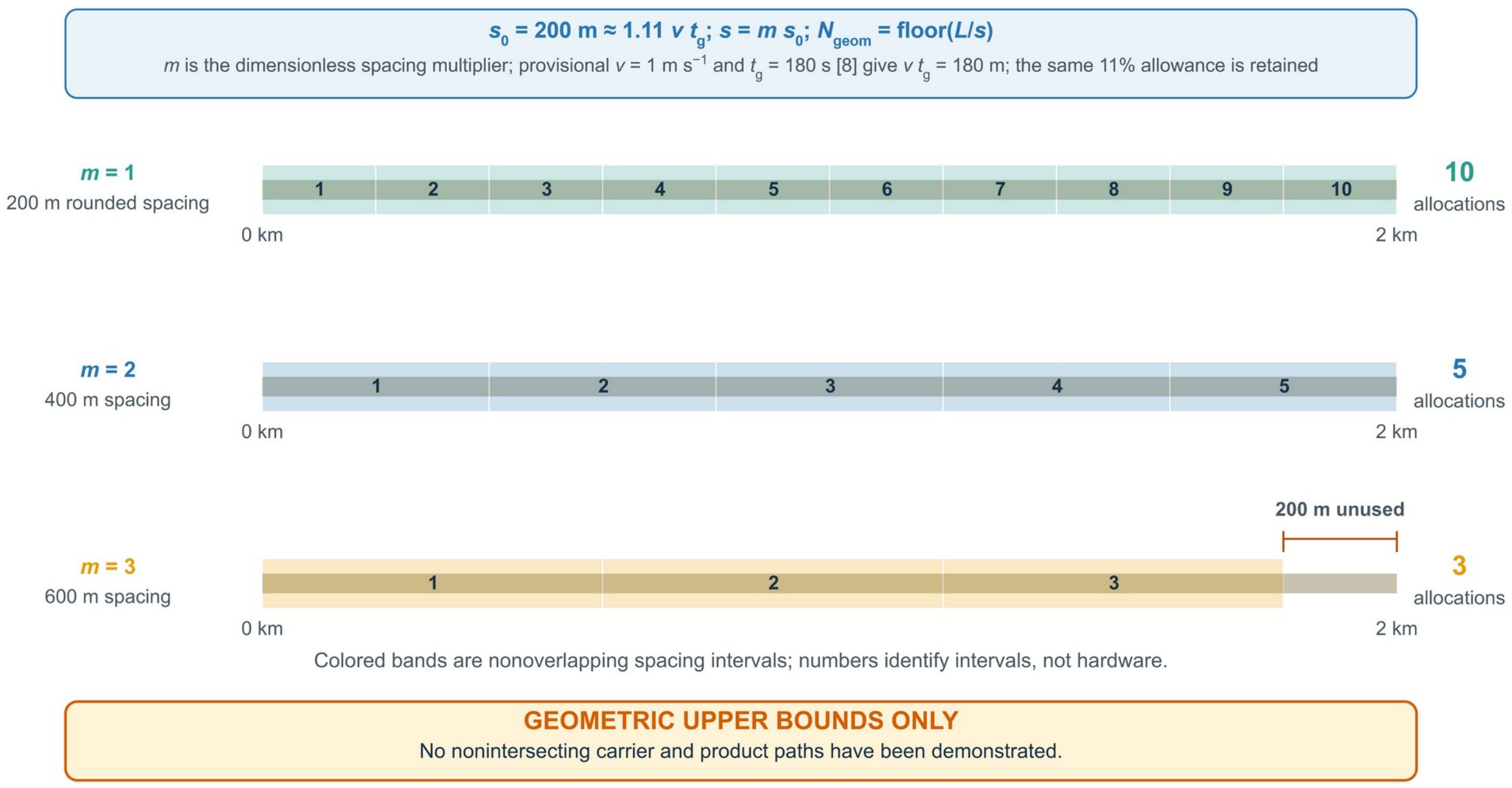

Figure S3. Illustrative geometric track-allocation upper bounds. For provisional inputs of $v$ = 1 m $s^{-1}$ and $t_g$ = 180 s,[8] $v\ t_g$ = 180 m. The rounded base allocation $s_0$ = 200 m ≈ 1.11 × $v\ t_g$ supplies an 11% allowance; the plotted spacing is $s = m\ s_0$, where $m$ is a dimensionless spacing multiplier. The colored bands mark nonoverlapping spacing intervals; numbers identify intervals, not hardware. The geometric count is $N_{geom}$ = floor($L/s$). Values $m$ = 2 and 3 give five and three allocations; the last case leaves 200 m unallocated. These counts do not represent production units because no nonintersecting carrier-and-product paths have been demonstrated.

Table S4. Quantitative-assumption and claim-evidence matrix for the integrated concept.

| Central claim | Status | Evidence that would close it |
|---|---|---|
| Provisional full-area growth in ~180 s and acceptable repeated complete cycles | Reference 8 identifies work in progress in our group that has yielded 180 s coverage; the result is not yet publicly available. Repeated complete-cycle performance is unmeasured. | Publish the work-in-progress conditions and mapping. Then repeat growth, dwell to the scheduled peel, removal, and regrowth. An accepted product at each peel directly qualifies the chosen cycle; use feed-on/feed-off diagnostics or gating only if needed. |
| Near-single-crystal product over a declared length or area | Aspirational, measurement-limited designation; no universal acceptable grain-boundary density is assigned. | State mapped area, resolution, and sampling. If no boundary is detected, report the detection limit. If boundaries are present, report their density and the grain-size distribution; application requirements decide acceptance. |
| $Ni_{80}Cu_{20}(111)$ preferred substrate | Extrapolated | Surface Cu/Ni/C/S vs time under process gas; segregation model |
| Graphene/Ni(111) interaction is purely covalent or purely physisorptive | Not supported; calculations give a method-dependent mixed picture | Retain both interaction terms; measure adsorption state and effective adhesion separately |
| Two close minima imply easy decoupling | Unsupported | Calculate the transition barrier and test hot peeling under no directed gas, Ar, and optional reactive-gas conditions; the energy difference between minima is not the transition barrier. |

| Central claim | Status | Evidence that would close it |
|---|---|---|
| Optional directed-gas transport at the peel line | Unmeasured; not required if no-directed-gas peeling succeeds | Test no directed gas first, then Ar, then $H_2$ or activated H. If gas must act over 10 mm at 1 m $s^{-1}$, demonstrate effective transport ≥~5 × $10^{-3}$ $m^2$ $s^{-1}$ and any required reaction within 10 ms, with no residue or damage. |
| Reactive-gas-decoupled state stable at the peel temperature | Analogy only; optional | Only if reactive gas improves peeling, demonstrate the driven state and measure etching, wake-gas recovery, regrowth delay, and film quality. Failure does not reject successful no-flow or Ar conditions. |
| Repeated alignment and merging of multiple nuclei | Established in suitable graphene growth systems,[9–11] but not demonstrated after repeated hot removal in the proposed process. | After each peel, map orientation, stitching, layer number, and any accumulated boundary or defect population. A retained physical seed is not required if repeated seamless merging is demonstrated. |
| Added peel-zone gas lowers effective adhesion without damage | Unsupported; optional | Measure G(v, T, atmosphere) over the 0.3–10 J $m^{-2}$ design envelope for no directed gas, Ar, and reactive-gas conditions; retain the simplest condition that gives stable, damage-free peeling and regrowth. |
| Stable 1 m $s^{-1}$ peel across 30 cm | Extrapolated | Width × speed ladder with defect statistics |
| Complete single-head cycle | Calculation basis only; complete product path unvalidated. | Demonstrate a complete product path and map carrier composition/thickness, modulus, creep, CTE mismatch, roller radius, tension, adhesion hierarchy, downstream release, contamination, and reuse. |
| Multiple-head operation | Geometric slot count only | Define and demonstrate nonintersecting carrier and product paths before crediting any multi-head output. |
| 2 km × 30 cm single-crystal M(111) | Unsupported | Long-duration hot soak with progressive width/length; map orientation, translational registry, roughness, support sag, creep/warp, axial and transverse temperature, pressure drop, reactant depletion, and local flatness. |
| hBN as a separate manufacturing case | Supported by monolayer, multilayer, thick-film, and reuse precedents; hot release and integrated repetition are unmeasured. | Choose target thickness and M(111), then map phase, B/N activity, polarity, boundaries, hot adhesion, fracture, repeated delamination and growth, apparatus, and scale. Do not require the graphene apparatus or module length. |

## S5. hBN-specific decision matrix

Graphene and hBN do not share one growth-and-release window, and I do not assume that they share one apparatus or scaling strategy. Cu(111), Cu–Ni(111), Ni(111), and liquid-mediated Ni–Si/Ni(111) precedents correspond to different products and process windows.[15–19, 22] Cu–Ni(111) is important for monolayer hBN, whereas Ni(111) has produced trilayer and much thicker single-crystal films. The binary hBN lattice adds polarity, inversion and antiphase boundaries, B/N activity, and boride formation. No cited study demonstrates rapid hot release. That is the central experiment proposed here. Table S5 therefore defines hBN-specific choices and tests without pretending that every future parameter can be foreseen.

| Issue | Evidence / risk | Required measurement | Pass / stop rule |
|---|---|---|---|
| Catalyst and temperature | Cu(111), Cu–Ni(111), Ni(111), and liquid-mediated Ni–Si/Ni(111) precedents correspond to different monolayer, multilayer, and thick-film products.[15–19, 22] | Choose the intended hBN thickness first; then map M(111), phase, surface composition, roughness, temperature, and melting or dewetting margin. | Pass only with a stable crystalline catalyst and no damaging melt or dewetting. |
| B and N activities | B uptake can form nickel borides; N supply and loss are coupled. | Use in situ or quenched surface/bulk analysis and a Ni–B–N material balance. | Stop if boride formation or composition drift cannot be bounded. |
| Orientation and sublattice | An aligned diffraction pattern can hide 180° inversion or antiphase boundaries. | Map polarity, B/N termination, rotational variants, inversion domains, and boundary density. | Pass only when no disallowed boundary is detected over the stated area and resolution. |

| Issue | Evidence / risk | Required measurement | Pass / stop rule |
|---|---|---|---|
| Layer number | Published monolayer, trilayer, and thicker hBN are distinct process windows.[15–19, 22] | Map thickness and stacking over the product area after each cycle. | Stop or redefine the product if target-layer yield is not stable. |
| Regeneration and reuse | Ten-cycle reuse was reported for centimeter-scale bulk Ni(111) with hBN, and three growth-bubbling-transfer cycles for a 500 nm Cu(111)/4-inch sapphire film with graphene; another hBN transfer consumed the alloy.[15, 18, 57] | Repeat growth/removal while tracking catalyst structure and product acceptance metrics. | Pass only if quality and catalyst state remain stationary for a declared cycle count. |
| Hot delamination | No published study demonstrates rapid hot release of the cited hBN films. This is the experiment proposed here, not a presumed precedent.[15–19, 22] | Begin with coupon-scale hot release and repeated growth on the selected M(111); measure effective adhesion, fracture, carrier capture, product quality, and catalyst state. Build the speed, width, and apparatus ladder from those results. | Failure excludes the defined repeated hot-release hBN route even if batch growth succeeds. |
| Product acceptance | Electrical insulation alone does not establish a single crystal. | State mapped area, resolution, and sampling. Report grain-size distribution and boundary density when boundaries are present; report nonboundary macroscopic defects separately above the declared detection threshold. | Near-single-crystal requires no detected disallowed boundary. Near-zero nonboundary macroscopic defects per meter requires no detected listed defect and a reported confidence bound. |
| Apparatus and scale | Graphene and hBN need not share a reactor, carrier, release condition, product path, or preferred module length. | Use the selected hBN product and coupon results to define its apparatus and progression in width and length. | Do not reject hBN because it diverges from the graphene design; reject only a defined hBN route that fails its own gates. |

Table S5. Separate decision matrix for SC-hBN growth, regrowth, and hot delamination. Bracketed numbers are main-text references.

## S6. Carrier and facility material-energy boundary

The main-text energy numbers are order-of-magnitude screens for one head, using information available now. The wall-loss range brackets simple insulation conduction plus openings and thermal bridges. At 1 $m\ s^{-1}$ productive and return speeds with 200 s end handling, 200, 500, and 2000 m modules have cycles of 600, 1200, and 4200 s; rounded gross outputs of about 9,000, 11,000, and 12,000 $m^2\ day^{-1}$; and carrier-roll masses of about 12, 30, and 120 kg. The corresponding wall-energy screens are roughly 1–5, 3–10, and 10–40 kWh per gross square meter. Four assumed 90% serial yields and assumed 50% overall equipment effectiveness give a conforming fraction of $0.9^4 \times 0.5 \approx 0.328$ and wall-only values of roughly 3–15, 9–30, and 30–120 kWh per conforming square meter. These are not forecasts or a life-cycle assessment. Experiments and testing should replace and refine them.

| Boundary element | Representative basis | Present estimate | Needed measurement / treatment |
|---|---|---|---|
| Interfacial separation | Gwv with $G = 1\ J\ m^{-2}$, $w = 0.30$ m, $v = 1\ m\ s^{-1}$ | ~0.3 W per head | Negligible compared with facility heat; measure effective G at process conditions. |
| Carrier web | 25 µm metal-like carrier screen; ~0.20 $kg\ m^{-2}$; cp ~0.5 $kJ\ kg^{-1}\ K^{-1}$; ΔT ~1100 K; 1 $m\ s^{-1}$; q'' = 100–200 $kW\ m^{-2}$ | ~120 kg roll; ~2.5 $t\ day^{-1}$; ~40 kW during traverse; ~19 kW cycle average; Lh ~0.55–1.1 m | Map composition, 10–25 µm thickness, modulus, creep, CTE mismatch, roller radius, line tension, adhesion hierarchy, release energy, contamination, and reuse; no material is selected. |
| Hot enclosure | 9000 $m^2$ wall area; k = 0.05–0.10 $W\ m^{-1}\ K^{-1}$; 0.10–0.20 m insulation; ΔT ~1000 K; bridges/openings added | Rounded screen: ~0.5–2 MW (200 m), ~1–5 MW (500 m), and ~5–20 MW (2000 m); corresponding wall energy is roughly 1–5, 3–10, and 10–40 kWh per gross $m^2$. | Measure effective perimeter, penetrations, support conduction, radiation shields, and module-length sensitivity. |

| Boundary element | Representative basis | Present estimate | Needed measurement / treatment |
|---|---|---|---|
| Conforming product | Four serial yields of 90% and 50% overall equipment effectiveness | Illustrative assumptions only: $0.9^4 \times 0.5 \approx 0.328$ conforming fraction; wall-only screens of roughly 3–15, 9–30, and 30–120 kWh per conforming m² for 200, 500, and 2000 m modules. | Replace assumed yields with measured coverage, crystal, layer, peel, carrier-release, and uptime data; use sensitivity ranges rather than one facility value. |
| Gas, pumping, and cooling | Representative process-gas screen; $H_2$ case uses $10^3$–$10^4$ standard L min$^{-1}$ heated through ~1000 K. Ar, mixed-gas, lower-flow, and no-directed-peel-gas cases remain open. | $H_2$ sensible heat ~0.02–0.2 MW before recovery; this is one possible burden, not a gas selection. Pumping, abatement, and cooling remain unbounded. | Measure flow, composition, pressure drop, heat recovery, purge demand, flammable-gas inventory when applicable, abatement, and cooling duty for the selected growth and peel conditions. |
| Start-up and substrate | Illustrative 2 km × 0.30 m × 40 µm Ni web: mass ~210 kg; heat capacity ~0.5 kJ kg$^{-1}$ K$^{-1}$; temperature rise ~1000 K. The actual metal thickness is not selected. | This illustrative metal-web basis gives ~0.1 GJ; enclosure refractory, supports, and repeated start-up frequency could dominate and are unbounded. | Report start-up frequency and amortize energy over conforming production. |
| Carrier life cycle | Carrier manufacture, recovery, cleaning, and loss | Outside the present numerical estimate | Include mass yield and embodied energy in any life-cycle claim. |

Table S6. Material-and-energy accounting boundary for the illustrative single-head case. Values are design screens; unquantified terms must be included before comparing facilities or making life-cycle claims.

## S7. Extended liquid-metal-route analysis

Liquid-mediated growth includes at least two physically different cases. In the NUS archived preprint, Yang et al. reported a 3–30-nm-thick, 5.08-cm single-crystal rBN film by solid-liquid-interface-mediated epitaxy on Ni(111)/sapphire.[19] They proposed a thin Ni–Si surface liquid that stores B and N while the underlying solid Ni(111) preserves epitaxial orientation. This is an important example of a liquid participating in thick single-crystal BN growth. Reactive molecular-dynamics work has also examined hBN growth from molten Ni solutions.[60] Neither result is equivalent to growth on a free liquid-metal bath. A free liquid surface removes the solid crystallographic template and requires a separate architecture. Liquid Cu is the more credible monolayer-graphene medium because its carbon solubility is low, whereas liquid Ni dissolves more carbon and has produced bulk graphite during cooling.[23, 49, 55] Quasi-monocrystalline graphene crystallization on a liquid Cu matrix has also been described.[61] Liquid Cu–Ni could tune carbon inventory. Many parameters remain unknown; the purpose of Table S7 is to order the first discriminating experiments, not to solve them all.

A free liquid surface is smooth on average but not crystallographically registered or mechanically quiet. Surface-normal atomic layering was measured for clean liquid Ga and In, with decay lengths of 5.8 ± 0.4 Å and 3.5 ± 0.6 Å, respectively. The remaining roughness was consistent with thermally excited capillary waves.[62, 63] Recommended surface tensions and densities give gravity–capillary crossover lengths of ~4.1 mm for Cu and ~4.8 mm for Ni.[64, 65] These estimates do not bound a moving meter-scale bath, for which Marangoni flow, buoyancy convection, vibration, oxygen, sulfur, bubbles, and slag must be controlled. Molten-Cu growth has reported ~1.5–3 Å roughness,[66] and capillary-wave-driven domain motion threatens long-range registry.[67]

For the free-liquid-bath route, direct hot lift-off and carrier capture should be tested before scale-up. Cooling or solidifying the metal before removal reintroduces contraction, step bunching, and fold formation; on Cu–Ni(111), folds appeared abruptly near 1030–1040 K, while 1000–1030 K yielded fold-free films.[11] The experiment must therefore close composition, surface motion, lift-off, carrier capture, and film crystallinity on a static bath before length or production rate is considered. Raman and AFM establish layer number, roughness, and defect proxies; they do not by themselves establish single crystallinity.

| Gate | Question | Measurement | Pass → proceed | Kill criterion |
|---|---|---|---|---|
| G0 — Architecture fit | Does free-liquid lift-off require an entirely separate facility? | Paper-level review of melt temperature, atmosphere, containment, carrier, lift-off, and product path. | A bounded static-bath test architecture is identified. | No safe, testable static-bath architecture can be defined. |
| G1 — Orientation and stitching | Can a free liquid yield the required graphene crystallinity? | After lift-off, use selected-area electron diffraction, dark-field TEM, and other boundary-sensitive mapping. Use EBSD only for solidified metal. | The declared crystallinity target is met at the stated detection limit. | Irreducible misorientation or boundaries redirect the route to textured products. |
| G2 — Layer control | Does the Ni / Cu–Ni segregation dial work? | Layer number & uniformity vs Ni fraction, carbon budget, cooling rate (Raman / AFM) | Reproducible 1-, 2-, n-layer selection at target uniformity | No controllable layering → drop the Ni route |
| G3 — Flatness / stability | Does the free liquid stay flat enough at temperature? | Capillary-wave / convection amplitude, evaporation & composition drift (static → meter bath) | Undulation and drift within the flatness / contamination budget | Undulations imprint beyond spec, or evaporation uncontrollable |
| G4 — Direct hot lift-off and capture | Can the film be lifted directly from the hot liquid and captured intact? | Attempt hot lift-off from a static bath. Map tears, wrinkles, layer number, roughness, crystallinity, carrier capture, and contamination. | Direct lift-off preserves the declared product criteria. | Required solidification or capture damage defeats the defined route. |
| G5 — Scale | Do orientation, flatness, containment, composition, and lift-off survive greater width and length? | Advance only after G0–G4 pass: static coupon → wider bath → moving short ribbon. | The measured windows persist as scale increases. | Loss of any passed gate stops further scale-up. |

Table S7. Ordered first tests for the free-liquid-metal route. The table does not attempt to specify every future parameter. Reference 19 is a distinct solid-liquid-interface precedent, not evidence for a free-liquid bath.

## Supporting References